\documentclass[superscriptaddress,reprint,aps]{revtex4-2}
\usepackage{url}
\usepackage{multirow}  % for merging rows in tables
\usepackage{booktabs}  % for the nicer table column header with the underline
\usepackage[table]{xcolor} % needed for the alternating gray rows in the table
\usepackage{amsmath}
\usepackage[utf8]{inputenc}
\usepackage{algorithm2e}
\usepackage{amssymb}
\usepackage{bm}
\usepackage{mathtools}
\usepackage{graphicx} % Required for inserting images
\graphicspath{{./figures/}}
\usepackage{siunitx} % physical units
\DeclareSIUnit\cal{cal}
\newcommand{\bx}{\boldsymbol{x}}
\usepackage{float}
\usepackage[colorlinks=true, allcolors=blue]{hyperref}
\usepackage{nameref}
\usepackage{caption}
\newcommand{\bifold}{BIFOLD - Berlin Institute for the Foundations of Learning and Data, Germany}
\newcommand{\mlgroup}{Machine Learning Group, Berlin Institute of Technology, 10587 Berlin, Germany}
\newcommand{\google}{Google Deepmind, Berlin, Germany}
\newcommand{\korea}{Department of Artificial Intelligence, Korea University, Seoul 136-713, Korea}
\newcommand{\maxplanck}{Max Planck Institute for Informatics, 66123 Saarbrücken, Germany}
\newcommand{\fu}{Department of Mathematics and Computer Science, Free University of Berlin, Germany}
\newcommand{\luxembourg}{Department of Physics and Materials Science, University of Luxembourg, L-1511 Luxembourg City, Luxembourg}

\begin{document}
\preprint{APS/657-LDR}

\title{Analyzing Atomic Interactions in Molecules as Learned by Neural Networks}

\author{Malte Esders}
\affiliation{\bifold}
\affiliation{\mlgroup}

\author{Thomas Schnake}
\altaffiliation{Contributed equally to this work}
\affiliation{\bifold}
\affiliation{\mlgroup}

\author{Jonas Lederer}
\altaffiliation{Contributed equally to this work}
\affiliation{\bifold}
\affiliation{\mlgroup}

\author{Adil Kabylda}
\affiliation{\luxembourg}

\author{Grégoire Montavon}
\affiliation{\fu}
\affiliation{\bifold}
\affiliation{\mlgroup}

\author{Alexandre Tkatchenko}
\email{alexandre.tkatchenko@uni.lu}
\affiliation{\luxembourg}

\author{Klaus-Robert M\"{u}ller}
\email{klaus-robert.mueller@tu-berlin.de}
\affiliation{\bifold}
\affiliation{\mlgroup}
\affiliation{\google}
\affiliation{\korea}
\affiliation{\maxplanck}

\begin{abstract}
  \noindent
  While machine learning (ML) models have been able to achieve unprecedented accuracies across various prediction tasks in quantum chemistry, it is now apparent that accuracy on a test set alone is not a guarantee for robust chemical modeling such as stable molecular dynamics (MD). To go beyond accuracy, we use explainable artificial intelligence (XAI) techniques to develop a general analysis framework for atomic interactions and apply it to the SchNet and PaiNN neural network models. We compare these interactions with a set of fundamental chemical principles to understand how well the models have learned the underlying physicochemical concepts from the data. We focus on the strength of the interactions for different atomic species, how predictions for intensive and extensive quantum molecular properties are made, and analyze the decay and many-body nature of the interactions with interatomic distance. Models that deviate too far from known physical principles produce unstable MD trajectories, even when they have very high energy and force prediction accuracy. We also suggest further improvements to the ML architectures to better account for the polynomial decay of atomic interactions.
\end{abstract}

\maketitle

\section{Introduction}
Methods for modeling atomistic systems range between computationally cheap but less precise (e.g., classical force fields), to computationally expensive but more precise (e.g., first-principles calculations based on Density Functional Theory (DFT), coupled-cluster method with single, double and triple excitations (CCSD(T)), or quantum Monte Carlo techniques (see e.g.~\cite{rupp2012fast,von2020exploring,schutt2020machine,keith2021combining,pfau2024accurate})). Machine Learning Force Fields (MLFFs) are an emerging technology that tries to favourably position itself by being computationally efficient while simultaneously approaching the more expensive methods in accuracy~\cite{unke2021machine}.

Due to the many-body nature of the Schrödinger equation, the computational cost of accurate \textit{ab-initio} methods grows extremely fast (exponentially or steeply polynomially) with the number of particles in a system~\cite{szabo2012modern, piela2006ideas}. Conversely, approximate methods with a lower computational cost inevitably need to ``cut corners'' and therefore may not adequately represent the full complexity of a system under study~\cite{sherrill2009assessment, piana2011robust}. As a result, numerous quantum-chemical approximation methods have been developed, each with its own trade-offs.
The usefulness of these methods lies in the detailed understanding of their limitations, allowing one to choose the most appropriate method for the task at hand.

Despite the vast potential of MLFFs, they may ultimately only become trusted once their strengths and weaknesses are similarly understood. For instance, a common problem of ML models is that they sometimes do not extrapolate well beyond their training domain~\cite{bengio2017deep}, and MLFFs are no exception. Although research into transferable models that are trained on well-curated datasets that broadly cover chemical space is ongoing \cite{ani1, aimnet1, ani2x, illarionov2023combining, kovacs2023mace, anstine2024aimnet2}, for the foreseeable future there likely will not be a one-size-fits-all model. This necessitates a deeper analysis of the underlying prediction strategy. The non-linear nature of complex ML models prohibits our understanding of how they form predictions, particularly when it comes to identifying potential shortcomings. The current study serves as a crucial step to address this issue: based on recent advances~\cite{montavon2018methods, samek2021explaining, schnake2022higher, xiong2022efficient, xiong2023relevant}, we present a method to uncover in detail the prediction strategies and learned representations of MLFFs. On the basis of four common chemical principles listed below, we examine to what extent they are embodied by learning models.

So far, research into MLFFs focuses mostly on developing new architectures that improve over the state-of-the-art in terms of the validation error. While this error is an important metric, it alone poorly predicts generalization performance. In particular, the stability of MD simulations with models of comparable force errors is vastly different \cite{stocker2022robust,fu2022forces,frank2024euclidean}. Notably, the atomic interaction analysis we propose, provides a better understanding of the learned representation and shows \textit{why} certain networks produce (un)stable MD trajectories.

\begin{figure*}%[ht]
    \centering
    \includegraphics[width=1.0\textwidth]{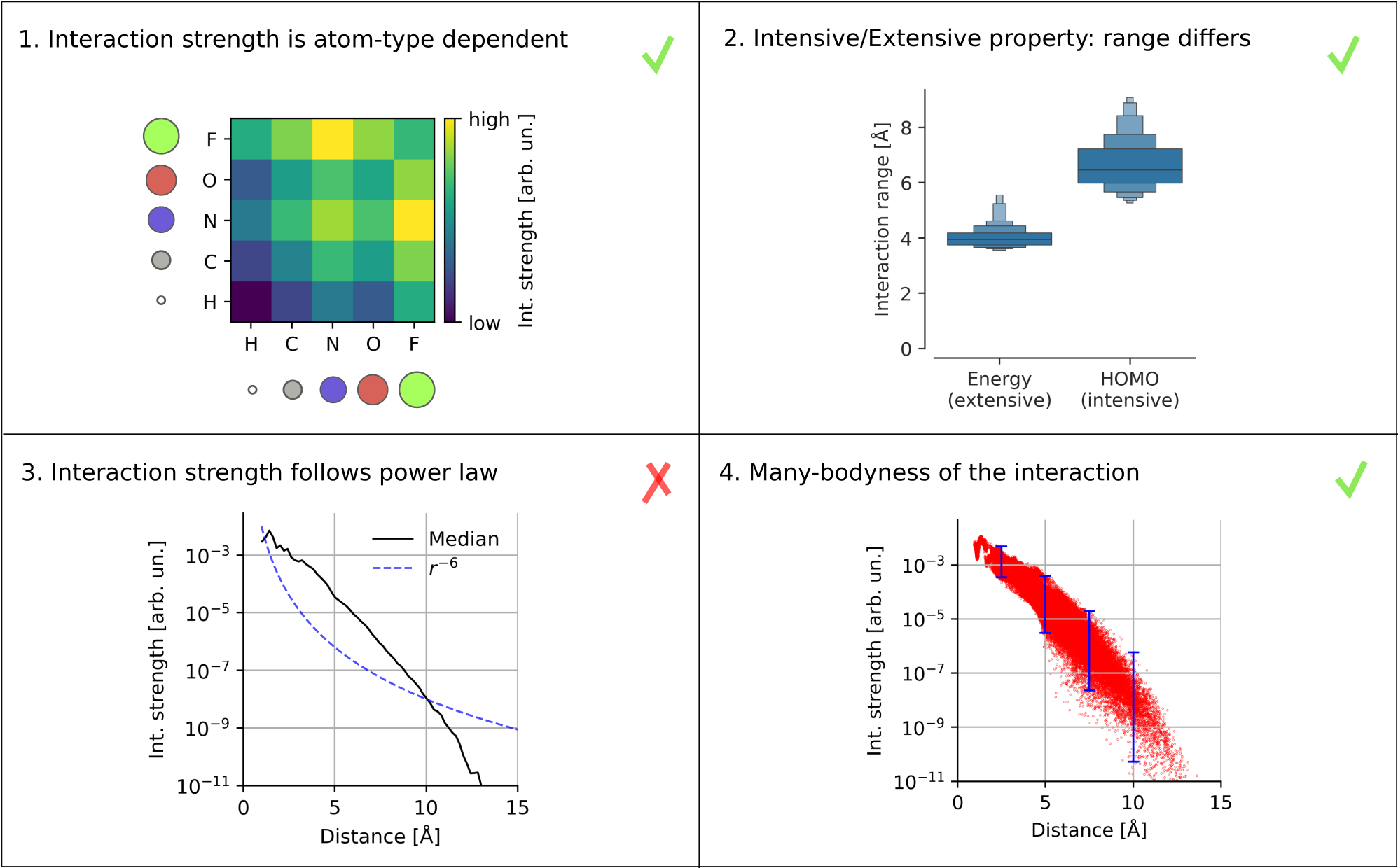}
    \caption{Using this study's explainability framework to inspect whether the models learned four common chemical principles from the data. For more details on the experimental setting, see Section \ref{sec:methodology}. Subfigure 1: Mean interaction strengths for atom-pairs at a distance less than $3$~\AA. The color-scale is logarithmic. Subfigure 2: Interaction range (eq.~\ref{eq:interaction_range_human_interpretable}) for a model trained on atomization energy (extensive property) and HOMO energy (intensive property) from the QM9 dataset. Subfigure 3: Median of the interaction strength across interatomic distance, provided by the XAI method, compared to $r^{-6}$, a typical decay for London dispersion forces (molecule: Ac-Ala3-NHMe). Subfigure 4: Spread of the interaction strength for different distances (each dot in the scatter plot is one atom pair); For selected distances, the maximum to minimum interaction strength (blue lines).
    }
    \label{fig:chemical_principles}
\end{figure*}

Training models is based purely on learning a mapping from atom positions and atomic numbers to properties like the atomization energy and the forces. It is generally hoped that models can learn the underlying physics purely from such data, but an analysis to which extent that is actually the case is so far lacking. In this study, we aim to fill this gap by proposing a way to systematically test the chemical plausibility of MLFF predictions. To this end, we posit the following four chemical principles:

\textbf{I. The relevance of interactions is atom-type and property dependent (found)}:
The relevance of atomic interactions predicted by MLFFs varies based on the atom types involved and the property being predicted. This atom-type and property dependence is particularly pronounced in bonded interactions, whereas at longer-range interactions, the dependence on the property becomes less pronounced.

\textbf{II. Different interaction range for intensive vs. extensive properties (found)}: Extensive properties can be approximated by evaluating the property on parts of the whole, and summing these local contributions up to obtain the property for the entire system~\cite{atkins2023atkins}. One could say the whole is the sum of the parts (at least up to a given accuracy). For intensive properties on the other hand, the entire system must be taken into consideration, and the whole is different from the sum of the parts. Therefore, one expects a higher interaction range when predicting intensive properties.

\textbf{III. Decrease of interaction strength with distance follows a power law (violated)}: At higher distance ranges, forces within molecules often fall off with a power law~\cite{stone2013theory}. For instance, forces between permanent dipoles fall off with $r^{-4}$, and London dispersion forces and dipole-induced dipoles fall off with $r^{-7}$ (when using the pairwise approximation).

\textbf{IV. Many-bodyness (found)}: The interaction strength should be anisotropic, meaning in this case that the interaction strength for equally distant atom pairs should differ depending on other atoms in the neighborhood~\cite{stone1988some, eramian2013anisotropy}. We call this property ``many-bodyness'', and contrast it with classical force fields, where interactions typically involve 4 or less directly bonded atoms. At higher distances, only 2-body terms are considered in widely used mechanistic force fields \cite{amber, charmm}.

An overview of these principles with some illustrative results can be found in Figure \ref{fig:chemical_principles}. We see in subfigure 1  that the interaction strength is atom-type dependent. Subfigure 2 shows that the extensive property of energy has a smaller interaction range than the intensive property of HOMO energy. Subfigure 3 shows that the median interaction strength at different atomic distances does not follow a power law, particularly it does not follow $r^{-6}$. Subfigure 4 shows that interaction strengths between atom pairs at the same distance differ, which is due to the effect of other atoms in the neighborhood, a phenomenon we call ``many-bodyness'' (see also Figure~\ref{fig:overview} bottom for an illustration).

While some of these chemical and physical properties might seem to be textbook knowledge, only qualitative guidelines can be formulated with our limited understanding of many-body quantum mechanics. In contrast, ML models learn a quantitative mapping between structures and QM properties within the chemical space defined by a given dataset. Hence, a natural and so far unanswered question is whether these \textit{quantitative} predictions also obey the known \textit{qualitative} chemical and physical principles. This is the main challenge addressed in the current work.

None of the discussed properties is given to the ML models as an inductive bias, i.e.,\ as an explicit part of their architecture or loss function; therefore, it is merely a hope that such principles will be learned from the data. In the current study, we test each of these properties on different MLFFs. Specifically, we show that the closer an MLFF agrees with the above principles, the more stable its MD trajectories are.

\subsection{Related work}
\textbf{XAI for quantum chemistry.}
Trying to analyze the prediction strategy of graph neural networks (GNNs) applied to molecular data started soon after using GNNs became popular in quantum chemistry. Early approaches analyzed the atom-wise energy contributions or introduced a test charge to measure the model's reaction \cite{schuett2019xaiqc,schutt2017quantum}. This approach is still in use today, for instance in assessing the robustness of the prediction \cite{chong2023robustness}.

Using first-order explanation methods like layer-wise relevance propagation \cite{lrp} can uncover which individual nodes are relevant to the prediction \cite{ying2019gnnexplainer,luo2020parameterized,cho2020layer}. Other approaches yield relevant clusters of atoms \cite{collins2023xaiMolFrag,el2024global,lederer2023automatic}. Such explanation approaches are useful for a variety of chemical applications \cite{jose20drugdisc,mcclo19usingatt}, but can not go beyond atomic or cluster relevances.

In \citet{schnake2022higher}, the authors provide a higher-order relevance attribution method for GNNs, which decomposes the model's prediction into contributions of atom sequences, referred to as ``walks''.
Higher-order relevance attributions can be associated with many-body interactions~\cite{schnake2024symbolicxaiexplanation} and have helped to corroborate the importance of such interactions in coarse-grained protein systems~\cite{bonneau2024peering}. We will use this method in our framework.

\textbf{Chemical plausibility of ML models.}
Recently, several studies highlighted the need to move beyond just the validation accuracy, because the validation accuracy was shown to be insufficient to predict MD stability. Therefore, it alone is not a good measure of the degree to which chemical principles were learned from the data~\cite{miksch2021strategies,stocker2022robust,fu2022forces,wang2023improving}.

The FFAST software package~\cite{fonseca2023force} is an example of a tool designed for detailed analysis of MLFF prediction results, including visualization of per-atom prediction errors, force errors densities, and challenging conformers. While such analysis can be invaluable, the current study aims to go beyond that by investigating the underlying GNN prediction strategy and understanding why prediction errors occur, rather than merely identifying whether and where they happen.

A particular focus of this study is the long-range interactions within molecules. For symmetric Gradient Domain Machine Learning (sGDML) global models~\cite{MD17} that do not impose any cutoff, it was found that atom interactions up to~\SI{15}{\AA} are crucial for maintaining high accuracy \cite{kabylda2023efficient}. It is important to note that the diameter of the studied molecules was roughly \SI{15}{\AA}, suggesting that relevant interactions might span larger distances in larger molecules. Additionally, it was found that many of these long-range interactions could be removed without losing accuracy, whereas most of the short-range interactions needed to be kept. This indicates that, at least in sGDML-type models, only some long-range interactions are relevant. The analysis presented in the current study provides a way to identify the most relevant interactions in GNNs as well.

\subsection{Overview of this study and its contributions}
We focus our analysis on GNNs as a popular implementation of MLFFs. We make use of a recently proposed XAI method, called GNN-LRP \cite{schnake2022higher}, that allows to assign a relevance to sequences of nodes in the graph. In a first step, we review GNNs for quantum chemistry and XAI, specifically the GNN-LRP method, and outline how this method can be used in quantum-chemical applications.

We then extend GNN-LRP specifically for MLFFs. Making use of the fact that molecular ``graphs'' are embedded in Euclidean space, we propose a distance measure for sequences of nodes and use it to compute the interaction range that a GNN uses to form its prediction. Additionally, we develop a measure for the interaction strength between atoms as seen by the GNN. Lastly, we propose a measure for the many-bodyness of the interaction strength.

We then apply these methods to the representative and popular SchNet \cite{schutt2018schnet} and PaiNN \cite{schutt2021equivariant} architectures in various atomistic settings. SchNet and PaiNN use rotationally invariant and equivariant message passing, respectively, which allows us to compare the prediction behavior of these fundamentally different architectures. We provide a detailed analysis of each of the four chemical principles stated above and whether they are expressed in the models. Additionally, we provide a way to go beyond the classic ``generalization error'' as a performance metric, and use our proposed analysis to predict the stability of MD-trajectories.

\section{Methodology}\label{sec:methodology}
\subsection{Graph Neural Networks for Quantum Chemistry}\label{sect:GNN}
Most state-of-the-art MLFFs~\cite{unke2021machine} are from the family of GNNs~\cite{zhou2020}. GNNs for quantum chemistry work in two phases. In the first phase, each atom, indexed by $k$, gets represented as a point in a high-dimensional ``feature space''. This is achieved by initializing the atoms to element-specific embeddings and then iterate $T$-times a ``message passing'' step between atoms within a certain cutoff distance, resulting for atom $k$ in a vector representation $\bm{H}_{T,k}$ after the $T$-th message passing step. After the feature representations are updated by several message-passing steps, they encode the local chemical environment of each atom and thus contain the relevant information about molecular geometry and composition. Then, in the second phase, a feed-forward neural network predicts molecular properties from the atomic feature representations.

SchNet~\cite{schutt_schnet:_2017, schutt2018schnet} and PaiNN~\cite{schutt2021equivariant} are variants of GNNs applied to 3D geometries. They derive a connectivity graph where the graph nodes represent the atoms and the graph edges describe to what extent neighboring atoms are directly interacting. The connectivity of the graph is determined by a cutoff distance, beyond which all direct connections between nodes (atoms) are cut. A ``cutoff-function'', usually a cosine \cite{behler2017}, is applied to the interactions to ensure that there is a smooth transition toward the cutoff. A single message passing step is represented by a so called interaction block. For the considered architectures, several interaction blocks are stacked to ensure that also distant nodes can exchange information, as well as to allow the nodes to build a more fine-grained embedding of their atomic neighborhood. While SchNet solely learns scalar feature representations in the first phase, PaiNN in addition learns vectorial features~\cite{schutt2021equivariant}. The rotational equivariant nature of those vectorial feature representations makes PaiNN more data efficient~\cite{batzner2021se} and, as a result, provides more stable MD trajectories~\cite{frank2024euclidean}.

The first phase of the GNN, the message passing step, can be further divided into two individual steps, the aggregation step and the combine step. In the \textit{aggregation} step the incoming ``messages'' from an atom's neighboring atoms are aggregated, and in the \textit{combine} step the aggregated messages are combined non-linearly with the respective atomic feature representation of the node. Hence, the GNN is of the form
\begin{align}\label{eq:GNN}
     \bm{H}_{t+1,k} = \mathcal{C}\left(\sum_{j \in \text{neigh}(k)} \mu(\bm{H}_{t,k}, \bm{H}_{t,j}, r_{kj})\right)
\end{align}
where $\mu$ and $\mathcal{C}$ are message and combine functions, respectively, and $r_{kj}$ is the distance between the atoms indexed by $j$ and $k$. The set $\text{neigh}(k)$ specifies the neighbors of atom $k$ that are within the cutoff distance. The sum over the messages of all neighboring atoms yields the aggregated message.

In the following, we denote for each atom, indexed by $i$, $\bm{r}_i \in \mathbb{R}^3$ to be its position. In addition, we consider $f: \mathcal{X} \rightarrow \mathbb{R}$ to be the ML model with a scalar prediction. The domain $\mathcal{X}$ of the model in our case is the set of all possible geometric configurations of atoms. Each molecule is represented by the positions $(\bm{r}_i)_i$ of its atoms, indexed by $i$, and their respective nuclear charges.

\subsection{Explainable AI}
ML models, in particular deep neural networks, have demonstrated high predictive capabilities for a broad range of tasks, including accurate inferences of molecular electronic properties in the field of quantum chemistry~\cite{schutt_schnet:_2017}. These models, while achieving high accuracy, are fundamentally black boxes. In other words, they do not achieve the objective of shedding light on the structure of the inferred input-output relation, which is a more fundamental scientific objective~\cite{hoffmann2020simulation}. Furthermore, the measured accuracy may conceal whether the learned relation is physically meaningful, or whether it arises from exploiting a confounder in the data, the so-called Clever Hans effect \cite{lapuschkin-ncomm19,samek2021explaining,anders2022finding,kauffmann2024clever}.

XAI (see e.g.~\cite{samek2021explaining}) is a recent trend in ML, which aims to gain transparency into these highly complex and powerful ML models. Through specific algorithms operating on the structure of the learned ML model,
XAI helps clarify the strategy an ML model uses to generate its predictions.
XAI has multiple applications: It enables, together with a human expert, to validate an ML model, in particular, detecting features that an ML model uses as part of a Clever Hans strategy \cite{lapuschkin-ncomm19} (aka.\ shortcut learning \cite{DBLP:journals/natmi/GeirhosJMZBBW20}). Another application of XAI is in serving as scientific assistants \cite{DBLP:journals/mima/ZednikB22,klauschen-pathology24}, where, alongside a well-trained ML model, it helps to identify candidate input-output relationships for further testing by human observers in subsequent targeted experiments.

The field and the set of proposed XAI methods is highly heterogeneous. This is partly due to the broad range of meanings of the terms such as `explainability' and `interpretability,' as well as the diversity of practical use cases. However, research has coalesced around specific problem formulations, one of which is the problem of attribution.

Attribution assumes an input domain $\mathcal{X}$, an output domain $\mathbb{R}$, typically real-valued, and a prediction function $f:\mathcal{X} \to \mathbb{R}$ linking instances in the input domain to values in the output domain. In a quantum chemistry context, the input can be a set of features describing the molecular geometry, and the output the electronic property (e.g.,\ atomization energy). Focusing on a single prediction $\bx \mapsto y$ with $\bx = (x_1,\dots,x_d) \in \mathcal{X}$ the collection of input features and $y = f(\bx) \in \mathbb{R}$ the real-valued output, we would like to compute for each input feature $x_i$ a score $R_i \in \mathbb{R}$ measuring the extent by which this feature has contributed to the output $y$. Many methods have been proposed to compute these scores, e.g.,\ \cite{DBLP:journals/jmlr/StrumbeljK10,DBLP:conf/icml/SundararajanTY17,lrp}, with different properties in terms of robustness, computational efficiency, and applicability. One such method, \textit{Integrated Gradients} \cite{DBLP:conf/icml/SundararajanTY17}, assumes that the function $f$ is differentiable and that the point $\bx$ of interest is linked to a root point $\widetilde{\bx}$ through some segment parameterized by $\alpha$, and decomposes the prediction $y$ in terms of input features via the equations:
$$
y = \int \frac{\partial y}{\partial x} \frac{\partial x}{\partial \alpha}  d \alpha
= \sum_{i=1}^d \underbrace{\int \frac{\partial y}{\partial x_i}  \frac{\partial x_i}{\partial \alpha} d\alpha}_{R_i}
$$
where for practical purposes the integral is discretized, typically into 10-100 steps. Variants of the equation above, involving multiple potentially nonlinear paths, are possible. For those path-based methods to work well, one should assume that the path remains on the data manifold, so that the model's behavior is evaluated on regions of the input space that are physically meaningful. In a quantum chemical scenario, where atomic coordinates or interatomic distances form the input representation, one may be required to define an appropriate path between the current molecule and some reference molecule (e.g.,\ a relaxation path). Such path may however be unknown, or there may be multiple ones.

An alternative approach to determine the scores $R_i$, which does not require defining a root point or an integration path, and which requires only one forward/backward pass into the network is \textit{Layer-wise Relevance Propagation} (LRP) \cite{lrp}. LRP leverages the structure of the ML model that has produced the prediction. In particular, it assumes the mapping from input to output is given sequentially by the multiple layers of a deep neural network, i.e.,\ $\bx \mapsto\dots \mapsto (H_j)_j \mapsto (H_k)_k \mapsto \dots \mapsto y$, where $(H_j)_j$ and $(H_k)_k$ denote the collection of activation in two consecutive layers. LRP starts at the output of the network, and decomposes the prediction $y$ to neurons in the layer below. These scores are then backpropagated from layer to layer, using purposely defined propagation rules, until the input features are reached. Let $(R_k)_k$ be the scores resulting from propagating from the top layer until the layer with neurons indexed by $k$. Propagation to neurons of the layer below can be achieved using a rule of the type
\begin{align}
R_j = \sum_k \frac{z_{jk}}{\sum_{j'} z_{j'k}} R_k \label{eq:lrp_classic}
\end{align}
where $z_{jk}$ quantifies the contribution of neuron $j$ to the activation of neuron $k$. The multiple ways the scores $z_{jk}$ are defined give rise to different LRP propagation rules (cf.\ \cite{DBLP:series/lncs/MontavonBLSM19} for an overview). Numerous instantiations of LRP have been proposed, covering models as diverse as convolutional neural networks \cite{lrp,DBLP:series/lncs/MontavonBLSM19}, LSTMs \cite{DBLP:series/lncs/ArrasAWMGMHS19}, transformers \cite{DBLP:conf/icml/AliSEMMW22}, classical unsupervised learning models \cite{DBLP:journals/pr/KauffmannMM20,DBLP:journals/tnn/KauffmannERMSM24}, and GNNs \cite{schnake2022higher}. Unlike methods based on integrated gradients, LRP benefits from the internal abstractions of the neural network. In the context of a quantum-chemical application, this allows to attribute the prediction in terms of atoms and their relative distances without having to define meaningful paths for the molecule in the input space.

We note that all explanation techniques we have described so far produce an attribution of the prediction onto individual features, which in our quantum chemical scenario could be atoms and interatomic distances. Note that for GNNs, as we treat here, a decomposition onto individual atoms is readily available from the GNN itself, because it predicts atomic contributions to the final predicted quantity. These explanations may provide useful insights into the model, but are strongly limited in their expressive power and their ability to generate useful hypotheses. For example, they do not say anything about whether the property of interest is the result of individual feature contributions (e.g.,\ localized atom-wise contributions), or whether it arises from the interactions of many of these features (e.g.,\ long chains of atoms spanning the whole molecule). To tackle this question, it is essential to move beyond classical attribution techniques and towards higher-order explanations, that are able to capture those more complex interactions.

\subsection{Higher-Order Explanations for GNNs}
\begin{figure*}%[ht]
    \centering\includegraphics[width=1.0\textwidth]{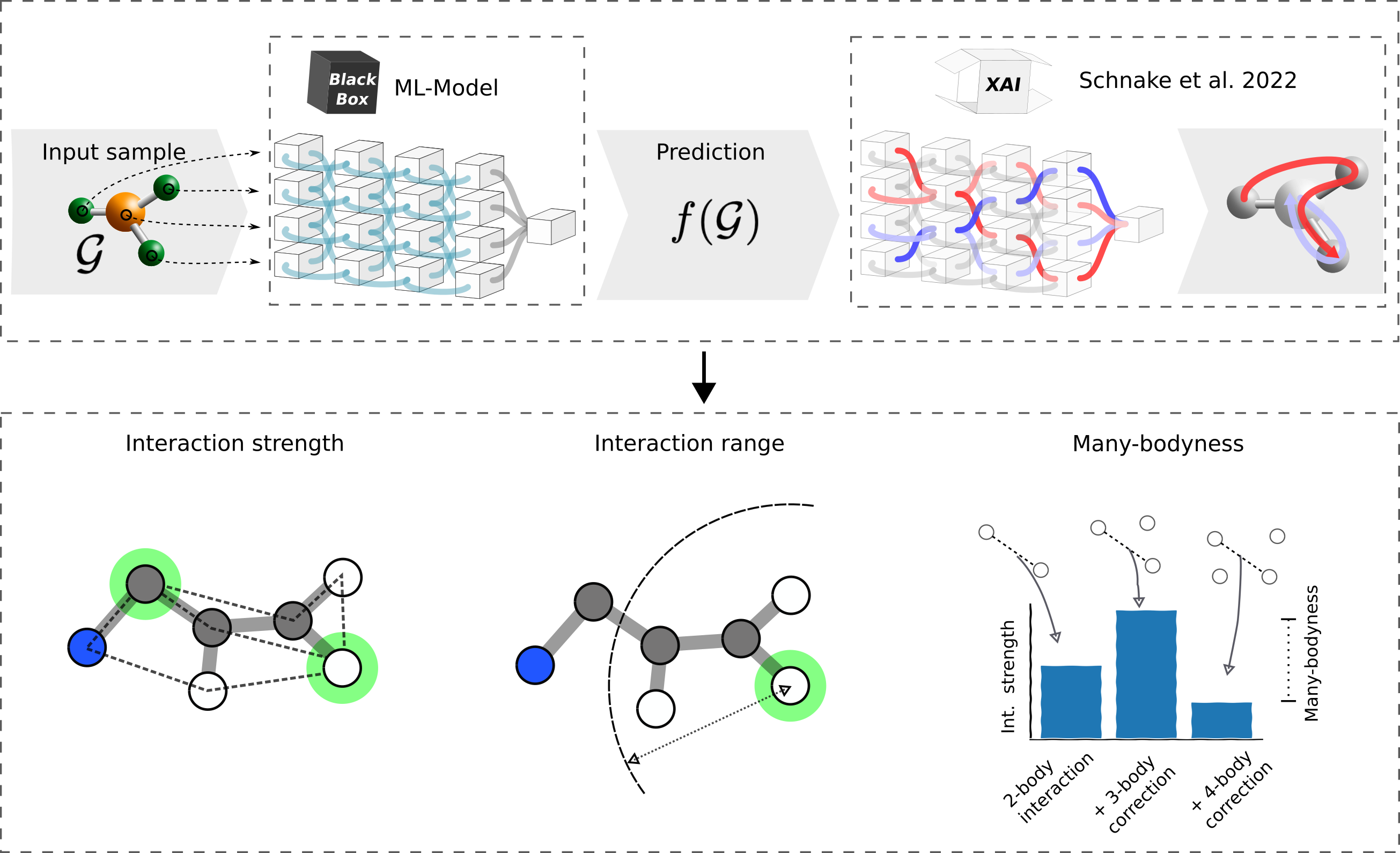}
    \caption{Overview of the explanation framework introduced in this study. A molecular input graph is processed by a black-box ML-model, specifically a GNN. The prediction is related to the input graph in the form of relevant walks on the graph, which are obtained from GNN-LRP \cite{schnake2022higher}. We extend this analysis to quantum chemistry-specific settings, using relevant walks on the graph: we provide a measure of the interaction strength between two atoms in a molecule (eq.~\ref{eq:interaction_strength}); we define the range up to which the network considers significant interactions (eq.~\ref{eq:interaction_range_human_interpretable}); and we specify the many-bodyness, which is a measure for how much the chemical neighborhood influences the interaction strength between two atoms (eq.~\ref{eq:manybodyness}).}
    \label{fig:overview}
\end{figure*}

Classical first-order attribution methods, as specified above, are limited to single feature attributions when predicting molecular properties. Even in simple scenarios, this approach is insufficient to understand the prediction strategy of the model. We believe that it is important to understand not only the relevance of each individual atom but also the nature and strength of the interactions between atoms from the model's perspective.

A seemingly straightforward approach to obtain interaction strengths would be to slightly perturb a given atom \textit{A} in various directions and record the change in atomic energy contribution at a target atom \textit{B}. One could then interpret the change in \textit{B}'s energy contribution as an indicator of the interaction strength from \textit{A} to \textit{B}, and vice versa. We caution that this approach is not as straightforward as it seems: A change in interatomic distance necessarily induces changes in other interatomic distances with respect to other atoms. Thus, we are left with the original problem of determining the true contributor of the observed energy change, not to mention the risk of moving outside the manifold of the data distribution. Changing the input to the model always carries this risk, which is a known problem for explainability methods \cite{dombrowski2024,blucher2024decoupling}.

Instead, we can proceed by extracting the contribution of interacting atoms directly from the structure of the MLFF model. This can be achieved in the context of GNN models by the GNN-LRP method \cite{schnake2022higher}, which we present below. We recall from Section \ref{sect:GNN}, that a GNN associates at each layer $t$ and for each node (atom) $j$ a representation $\bm{H}_{t,j}$, which we abbreviate in the following as $\bm{H}_j$.

A naive application of LRP to this architecture would start at the output and redistribute the predicted value backwards, traversing the multiple atom representations at each layer. The procedure would stop when the first layer is reached, where relevance scores can be mapped to atoms according to their representation in the first layer. This procedure, however, does not account for the way the different atoms have exchanged messages in the higher layers. GNN-LRP addresses this shortcoming by recording the path that relevance propagation messages have taken, and this is achieved by applying a slight modification to eq.\ \ref{eq:lrp_classic}:
\begin{align}
    R_{jkl\dots} = \frac{z_{jk}}{\sum_{j'}z_{j'k}}  R_{kl\dots}
    \label{eq:gnnlrp-simple}
\end{align}
In other words, we strip the pooling operation $\sum_k$ and retain the index $k$ in the propagated relevance score. Propagating through all layers of the GNN, we end up accumulating more and more indices, resulting in relevance scores over sequences of nodes (referred to as `walks' $\mathcal{W}$). These walks are of length $T+1$, where $T$ is the depth of the GNN. So far, for the simplicity of the presentation, we have assumed that each atom is represented by one neuron. However, in real GNNs, it is represented by $m$ neurons, meaning $\bm{H}_{j} = (H_{j}^b)_{b=1}^m$. Taking this into account, we need to extend eq.\ \ref{eq:gnnlrp-simple} as:
\begin{align}\label{eq:gnnlrp}
    R_{jkl\dots}^b = \sum_{c} \frac{z_{jk}^{bc} }{\sum_{b',j'}z_{j'k}^{b'c}}  R_{kl\dots}^c
\end{align}
where $b$ denotes a neuron associated to node $j$, $c$ denotes a neuron associated to node $k$, and $z_{jk}^{bc}$ quantifies the contribution of the neuron $b$ in node $j$ to the activation of neuron $c$ in node $k$.

A visual description of the method, along with an explanation of how it is used to quantify the model's physical properties, can be seen in Figure \ref{fig:overview}. The GNN-LRP method is theoretically founded in the higher-order Taylor decomposition of the model's prediction and can be seen as a generalization of LRP \cite{lrp} and deep Taylor decomposition \cite{montavon-pr17}. Furthermore, as shown in \cite{schnake2022higher}, it satisfies the axiom of conservation, namely
\begin{align}
\sum_\mathcal{W}  R_\mathcal{W} = y,
\end{align}
where $y$ is the predicted value at the output of the GNN. The latter allows us to view the GNN-LRP explanation as a decomposition of the GNN output (e.g.,\ predicted molecular energy) in terms of all the walks $\mathcal{W}$ on the molecular graph. The complexity of the explanation method increases exponentially with the number of layers, however, there are ways to  lessen the computational complexity from exponential to polynomial \cite{xiong2022efficient,xiong2023relevant}.

\subsection{Walk-importance and walk-distance}
In the following, we describe how we use the walk-relevances obtained from GNN-LRP \cite{schnake2022higher} to evaluate different properties of the model and its prediction strategy (for an algorithm, see Section~\ref{sec:appendix:methods}). One quantity we will use throughout is the measure of \textit{importance} for a walk $\mathcal{W}$ which we define by
\begin{eqnarray}
    \mathbb{P}(\mathcal{W})= \frac{1}{Z} | R_\mathcal{W} |
    \label{eq:prob}
\end{eqnarray}
where $Z = \sum_{\mathcal{W}} |R_\mathcal{W}|$. Note that $\mathbb{P}(\mathcal{W})$ is a probability distribution of $\mathcal{W}$, i.e., $\mathbb{P}(\mathcal{W})$ has values between 0 and 1, and $\sum_{\mathcal{W} \in \Omega} \mathbb{P}(\mathcal{W}) = 1$, where $\Omega$ is the set of all walks for a given atomistic system.

One of the questions we are interested in  is how long the range of interactions between atoms, as seen by the model, are. In particular, for any higher-order message $\mathcal{W}$ we can consider some distance $d(\mathcal{W})$ that a walk $\mathcal{W}$ traverses on the molecule.
One natural option for such a distance measure is the diameter of the smallest sphere that encloses all atoms in the walk $\mathcal{W}$. This is given by
\begin{eqnarray}\label{eq:dist_def}
    d(\mathcal{W}) := \max_{i,j \in \mathcal{W}} || \bm{r}_i - \bm{r}_j ||
\end{eqnarray}
where $|| \cdot ||$ denotes the Euclidean norm. In the remainder of this text, we use this distance measure, for example, when we develop more advanced concepts like the interaction range of an MLFF.

\subsection{Interaction range}
An important factor to evaluate is the distance at which atoms still have a significant influence on one another. Although short-range interactions, particularly those between directly bonded atoms, dominate the total energy of a molecule, it is the long-range interactions that, despite their small magnitude, are responsible for interesting macroscopic behavior like protein folding \cite{sherrill2009assessment,tkatchenko2009accurate,ambrosetti2016wavelike}.

However, modeling long-range interactions in MLFFs also brings a significant computational cost, as the number of interacting atoms scales roughly cubically with distance (due to the increasing volume of the cutoff sphere). For these reasons, it is crucial to get a sense of the range of interaction which the model still takes into account.

We propose to measure interaction range by looking at the maximum distance among walks that are important, i.e.,\ not assigned a non-negligible probability, as measured by eq.\ \ref{eq:prob}. To this end, we set a probability threshold $p_{\mathrm{min}} = 0.001 \max_{\mathcal{W} \in \Omega} \mathbb{P}(\mathcal{W})$ based on which we can search for a walk with maximum distance:
\begin{align}
\lambda^\text{prob}_{0.001} = \max_{\{\mathcal{W}|\mathbb{P}(\mathcal{W}) \geq p_\mathrm{min}\}}  d(\mathcal{W})
\label{eq:interaction_range_human_interpretable}
\end{align}
Note that not including an importance threshold, or setting it to zero, would be akin to always return the theoretically maximum walk length, which is independent of the solution learned by the GNN model.

As an alternative measure of interaction range, we consider a high-order statistic of the distribution of walk lengths. A simple such statistic, that retains a distance-based interpretation, is the `generalized expectation':
\begin{align}
\lambda^\text{pow}_a = \Big(\mathbb{E}_{\mathcal{W} \sim \mathbb{P}}[d(\mathcal{W})^a ]\Big)^{1/a}
\label{eq:generalized_expectation}
\end{align}
where $a$ is a parameter. Setting $a=1$ corresponds to measuring the expected distance, and $a=\infty$ the maximum distance. With the same aim of focusing on large distances, but discarding negligibly probable ones, we opt for the value $a=4$ in our experiments, which is closely related to the kurtosis commonly used to model peaks in a data distribution.

Unless otherwise noted, in all figures in this article, the threshold-based measure defined in eq.\ \ref{eq:interaction_range_human_interpretable} is used. We consider both measures valuable, and to show that the conclusions drawn in this paper are not dependent on the choice of range measure, both measures are reported for all experiments in Table\ \ref{table:all_statistics_short}.

\subsection{Attributing Atom Interaction Strength}
We now want to consider the strength of interaction between two atoms $i$ and $j$. Chemically, the interaction strength between two atoms in a molecule is not well defined. The presence of other atoms in the neighborhood and the resulting many-body behavior makes it impossible to measure the 2-body interaction strength in isolation. Nevertheless, multiple approaches to measure the interaction strength exist. For example, the Laplacian of the electron density at a critical point along the bond path can be seen as correlating with the interaction strength \cite{richard1990atoms}.

In this study, we develop a new measure for the interaction strength as seen by a GNN. We focus on two different approaches. In the first approach we want to consider all possible walks $\mathcal{W}$ that traverse the atoms $i$ and $j$, but can also traverse other atoms in the molecule. We call this the \emph{inclusive} interaction strength, because it is incorporating the context of the interacting atoms as well. We define this interaction strength by
\begin{eqnarray}
    s_{ij}^\mathrm{incl} := \sum_{\{ \mathcal{W} | i \in \mathcal{W} \wedge j \in \mathcal{W} \} } \mathbb{P}(\mathcal{W})
    \label{eq:interaction_strength}
\end{eqnarray}
Another approach to measure the interaction strength would be to consider all walks $\mathcal{W}$ that contain only the atoms $i$ and $j$. In other words, it consists \emph{exclusively} of walk contributions corresponding to interactions between $i$ and $j$, without the incorporation of the surrounding atoms. Formally, this can be given by
\begin{eqnarray*}
    s_{ij}^\mathrm{excl} := \sum_{\{ \mathcal{W} | \text{set}( \mathcal{W}) = \{i, j\} \} } \mathbb{P}(\mathcal{W})
\end{eqnarray*}
where $\text{set}(\mathcal{W})$ is the set of atom indices in $\mathcal{W}$.

We decided that it is generally more important to measure the interaction strength of two atoms in the context of their surrounding, therefore we use the inclusive measure $s_{ij}^\mathrm{incl}$ in the remainder of this text.

\subsection{Measuring many-bodyness}
\label{methods:many-bodyness}
We refer to many-bodyness as the property where the interaction strength between two atoms is influenced by other atoms in the neighborhood. In the context of MLFFs, measuring many-bodyness is of particular interest because it highlights a fundamental difference from mechanistic force fields. To illustrate this, assume a simplified force field based on a two-body expansion. In this case, the atom-atom interaction energy is fully isotropic: no matter where in the molecule the two atoms are positioned, the energy contribution will always be the same. Even real-world force-fields that do use higher-order terms usually do not go above 4-body terms. And these 4-body terms are only among chains of covalently bonded atoms. Atom pairs at higher (non-bonded) distances are modeled with 2-body terms only in mechanistic force fields \cite{amber, charmm}.

This is incompatible with physical reality: the atomic neighborhood that atoms are embedded in plays a fundamental role in their interaction. The promise of MLFFs is that they learn to capture this many-body nature better, but it has yet to be shown to which amount this is actually the case.

We propose a definition for the many-bodyness of atom-atom interactions within a molecule. We would like to express by how many orders of magnitude the interaction strength differs for equally distant pairs of atoms. We define
\begin{eqnarray}
  \{s\}_R := \big\{s_{ij} \,\big| \, || \bm{r}_{i} - \bm{r}_{j} || = R \big\}
\end{eqnarray}
to be the set of atom-atom interaction strengths for which the distance is $R$. This formulation is based on the continuous distribution of distances. In practice, where we have limited amounts of data, the condition of equality has to be relaxed to approximate equality: $|| \bm{r}_{i} - \bm{r}_{j} || \approx R$. In other words, we are distributing the atom pairs into bins along the interatomic distance. Then, we define the many-bodyness as
\begin{eqnarray}
  \gamma := \frac{1}{R_{max}} \int_0^{R_{\text{max}}} \log_{10} \Bigg(\frac{P_{100}(\{s\}_R)}{P_{10}(\{s\}_R)}\Bigg) \, dR
  \label{eq:manybodyness}
\end{eqnarray}
where $P_{100}(.)$ and $P_{10}(.)$ are percentile functions that return the 100th and 10th percentile. We use the 10th instead of the 0th percentile to be less sensitive to outliers. Using base $10$ for the logarithm instead of $e$ is chosen such that the resulting quantity can be interpreted as orders of magnitude.

Note that this measure of many-bodyness would be equal to $0$ for 2-body classical force fields (beyond the bonded cutoff distance), because all atom pairs of same elements at equal distance lead to the same energy contribution.

\begin{figure*}
    \centering
    \includegraphics[width=0.999\textwidth]{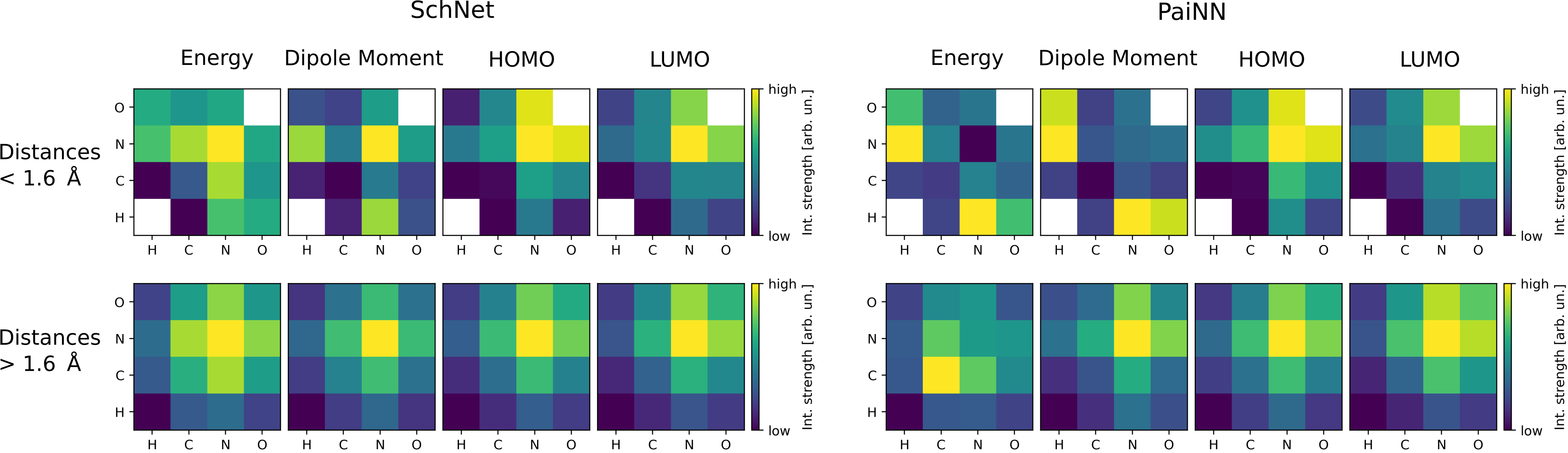}
    \caption{Mean log interaction strength for pairs of elements separated into `bonded' ($<1.6$~\AA), `non-bonded' ($>1.6$~\AA), and different quantum chemical properties. The networks are 3-layer SchNet and PaiNN trained on the QM9 dataset.}
    \label{fig:elem_interaction_strength}
\end{figure*}

\section{Results}

To analyze the first two chemical principles (i) relevance of interactions is atom-type and property dependent, and (ii) intensive properties require larger interaction range than extensive properties, we train SchNet and PaiNN models on four properties of the QM9 dataset. The properties are atomization energy, dipole moment, and highest occupied molecular orbital (HOMO) energies and lowest unoccupied molecular orbital (LUMO) energies. For the other two chemical principles (iii)-(iv), we trained SchNet and PaiNN on the molecule Ac-Ala3-NHMe from the MD22 dataset. The models were trained using SchNetPack~\cite{schutt2019schnetpack, schutt2023schnetpack}. For details of how the networks in this study were trained, cf.~ Section~\ref{sec:training_details}.

\subsection{Chemical principle 1: The relevance of interactions is atom-type and property dependent}

The nature of chemical interactions is fundamentally tied to the electronic configurations and corresponding atomic numbers of the elements involved. Within MLFFs, atomic numbers are encoded per-atom during training, resulting in learned interaction strengths that differ for each atom type, as expected. Figure~\ref{fig:elem_interaction_strength} illustrates the averaged interaction strength between four pairs of elements in the QM9 dataset (fluorine is omitted due to its low occurrence, with only 3 molecules containing it in the subset of the data). The atom pairs for four models each of the SchNet and PaiNN architectures are categorized into two length-scales: bonded ($< 1.6$\ \AA) and non-bonded ($> 1.6$\ \AA) interactions. Examination of these matrices reveals a clear atom-type and property dependence that the models have captured during training.

Interestingly, the interaction patterns differ significantly when comparing models within one architecture trained on different properties, and, for some properties, are drastically different when comparing the relevance between two architectures. This disparity is particularly evident when analyzing the 'bonded` interactions occurring within the 0.5-1.6\ \AA \ range (first row in Figure~\ref{fig:elem_interaction_strength}). For instance, while Nitrogen-Nitrogen interactions are deemed to be the strongest for energy prediction in SchNet, they are the weakest in the PaiNN architecture. Although we cannot definitively assert which representation is more accurate, it is reasonable to assume that a correct quantum projection exists for an atom-centered molecular basis representation~\cite{gallegos_explainable_2024,gori2023second}. This discrepancy underscores the importance of employing explainable artificial intelligence (XAI) techniques to analyze and interpret these complex relationships.

As we extend our examination to non-bonded interactions, we observe that the interaction patterns become more consistent across all properties and architectures. The atom-type dependence is visually preserved; however, it is now more closely related to the average distance between atom pairs. The observed decay of interaction strength with distance will be analyzed in more detail within Section~\ref{sec:results:interaction_strength}.

\subsection{Chemical principle 2: Intensive properties require larger interaction range than extensive properties}
The QM9 dataset provides several quantum chemical properties, of which some are extensive and some are intensive. Extensive properties can be thought of as ``the whole is the sum of the parts'', i.e.,\ additive local contributions sum towards the final quantity. Intensive properties are the opposite, where the quantity can only be determined by taking into consideration the entire molecule.

\begin{figure*}
    \centering
    \includegraphics[width=0.95\textwidth]{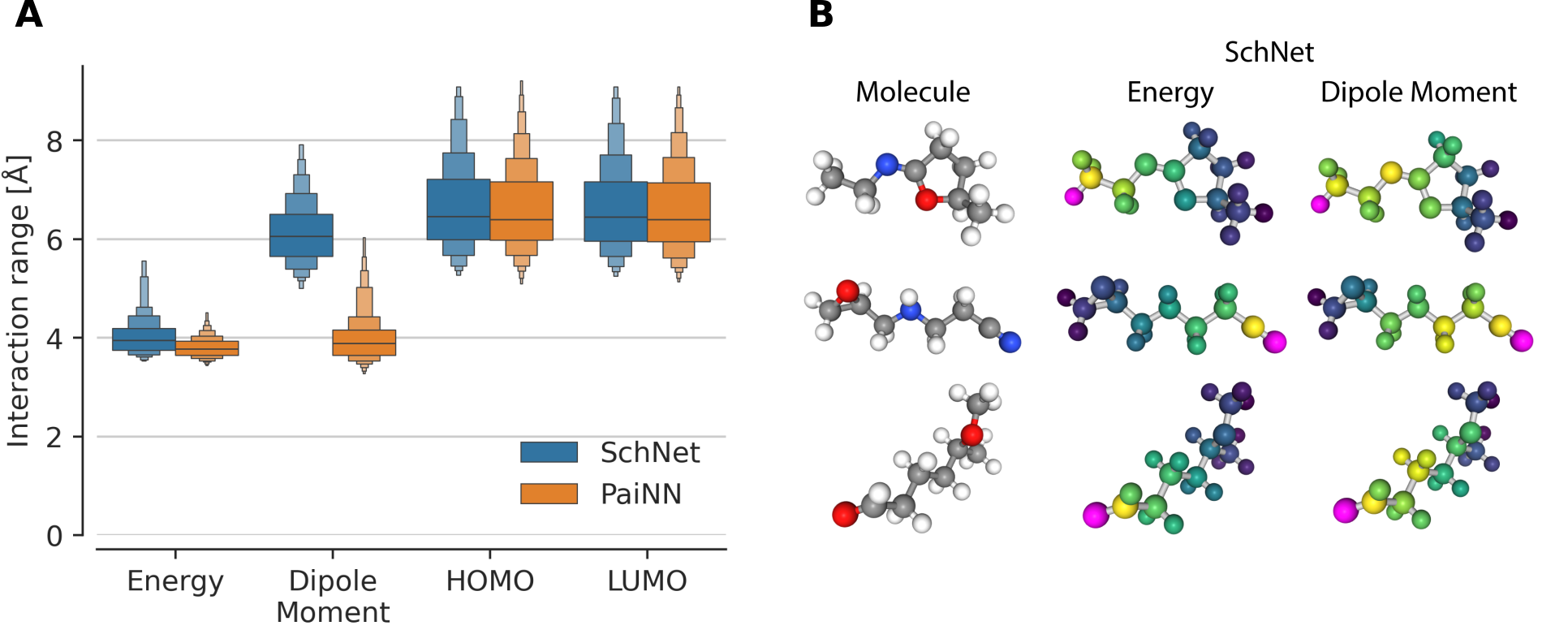}
    \caption{\textbf{Interaction range. (A)} SchNet and PaiNN calculated interaction ranges (eq.~\ref{eq:interaction_range_human_interpretable}) for models trained on various properties of the QM9 dataset. Energy is an extensive property, while dipole moment, HOMO and LUMO are intensive properties. \textbf{(B)} Examples of interaction strength for several molecules from QM9. The chosen atom is highlighted in purple, and the color scale ranges from dark blue (indicating weak interactions) to yellow (indicating strong interactions).}
    \label{fig:boxplot-extensive-intensive}
\end{figure*}

The extensive property we considered is the atomization energy. The intensive properties were the dipole moment, HOMO and LUMO energies. As described above, separate models of SchNet and PaiNN were trained for each property. The interaction range was determined with formula~\ref{eq:interaction_range_human_interpretable}. Figure~\ref{fig:boxplot-extensive-intensive}A shows that training on the intensive properties causes the models to learn to use a longer interaction range than the extensive properties, which is what we expected. Figure~\ref{fig:boxplot-extensive-intensive}B illustrates a difference of 2\ \AA \ in the interaction range of SchNet models trained on energies and dipole moments for three molecules from the QM9 dataset.

The interaction ranges learned by SchNet and PaiNN are remarkably similar for three properties. Keep in mind that SchNet and PaiNN have fundamentally different architectures, with SchNet being rotation invariant, and PaiNN rotation equivariant. The only property in which they diverge is the dipole moment: SchNet correctly treats it as an intensive property, while PaiNN is more expressive and might be constructing an internal representation of molecular dipoles as arising from interactions between local charges and dipoles.

\subsection{MD stability as an additional performance measure}
\label{sec:md_stability}
Datasets used for benchmarking MLFFs often come from single trajectory MD simulations~\cite{MD17, MD22}. In such datasets, the vast majority of samples are drawn from a small set of metastable states (local minima of the potential energy surface). As a result, state-of-the-art MLFFs achieve very low test errors, but this accuracy stems from the fact that the network has a rather ``easy'' interpolation task where many conformations are close to the energy minima. As an additional performance measure, it has been proposed to perform MD simulations with the MLFF and count how many simulations are unstable~\cite{fu2022forces,frank2024euclidean}.

\begin{figure*}
    \centering
    \includegraphics[width=\textwidth]{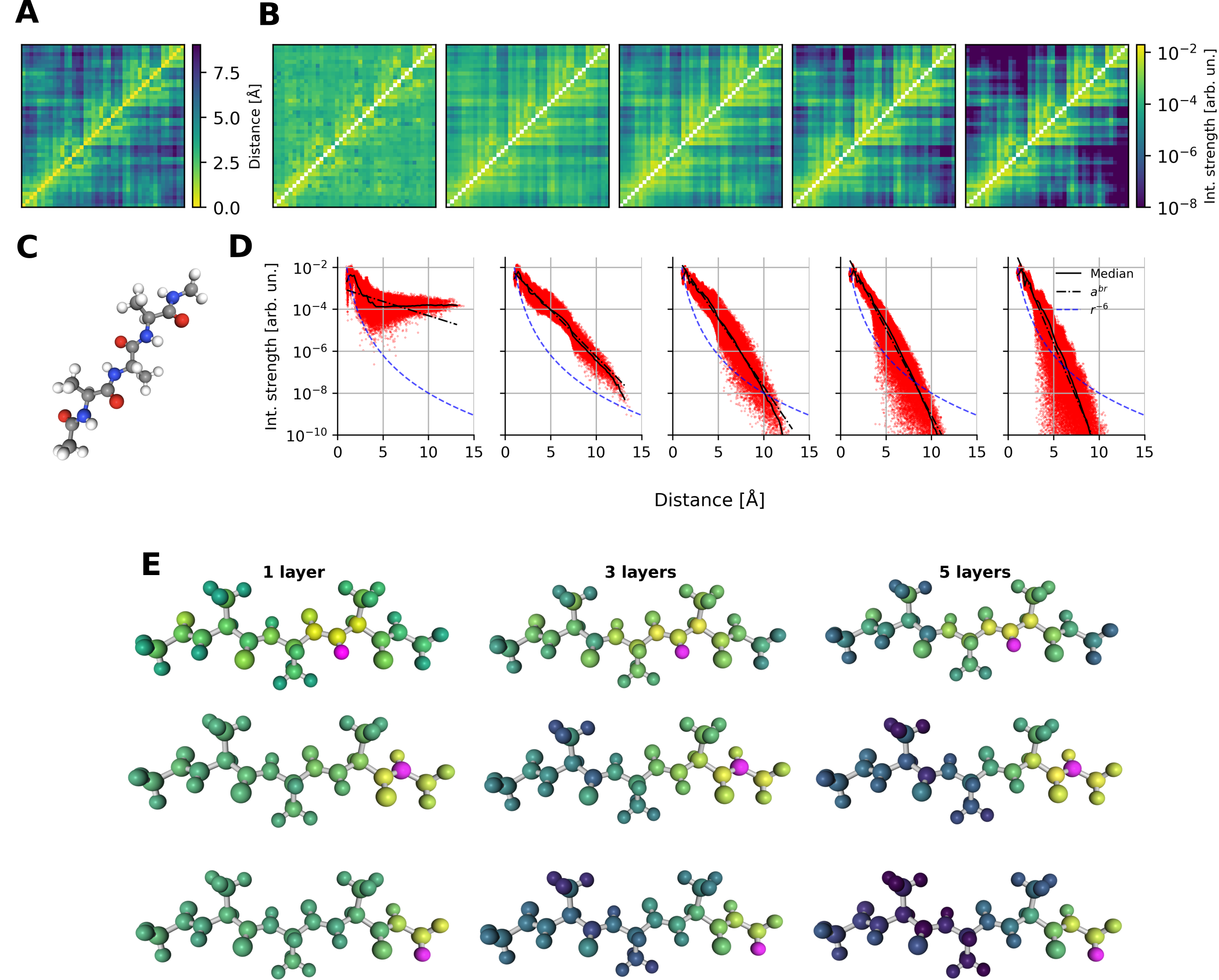}
    \caption{\textbf{Interaction strength of atom pairs in the tetrapeptide Ac-Ala3-NHMe. (A)} Distance matrix of one conformation of Ac-Ala3-NHMe. \textbf{(B)}: Atomic interaction strengths for 1 to 5-layer PaiNNs. All matrices were computed for the same randomly chosen conformer. \textbf{(C)} Structure of Ac-Ala3-NHMe. \textbf{(D)} Atomic interaction strengths as a function of distances for PaiNNs with 1 to 5 interaction layers, evaluated on Ac-Ala3-NHMe from MD22. The black lines indicate the median and the best fit of an exponential function, while the blue dashed line represents decay with $r^6$, a common model for London dispersion decay. \textbf{(E)} Examples of interaction strength for chosen (purple) atoms averaged over 100 conformations. The colorscale is consistent with B. }
    \label{fig:decay_per_layer}
\end{figure*}

All following experiments were computed from networks trained on the Ac-Ala3-NHMe molecule (Figure~\ref{fig:decay_per_layer}C) from the MD22 dataset~\cite{MD22}. Ac-Ala3-NHMe is a tetrapeptide containing 42 atoms and can exist in a folded and unfolded state, which makes it particularly interesting to analyze.

\begin{table}[]
  \centering
  \begin{tabular}{c|c|c|c|l|c|c|c}
    \textbf{$N_\mathrm{L}$} & Cutoff & $N_\mathrm{rbf}$ & $N_\mathrm{emb.}$  & Property & RMSE & MAE & MD failures \\
      \hline
      \multirow{2}{*}{1} & \multirow{2}{*}{15}   & \multirow{2}{*}{60}   & \multirow{2}{*}{512}  & Energy & 0.13 & 0.10 &  \multirow{2}{*}{22/30} \\
                         &                           &                       &                       & Forces & 0.25 & 0.19 & \\
      \hline
      \multirow{2}{*}{2} & \multirow{2}{*}{7.5}  & \multirow{2}{*}{30}   & \multirow{2}{*}{157}  & Energy & 0.17 & 0.13 &  \multirow{2}{*}{0/30} \\
                         &                           &                       &                       & Forces & 0.25 & 0.18 & \\
      \hline
      \multirow{2}{*}{3} & \multirow{2}{*}{5}    & \multirow{2}{*}{20}   & \multirow{2}{*}{128}  & Energy & 0.11 & 0.09 &  \multirow{2}{*}{0/30} \\
                         &                           &                       &                       & Forces & 0.14 & 0.10 & \\
      \hline
      \multirow{2}{*}{4} & \multirow{2}{*}{3.75} & \multirow{2}{*}{15}   & \multirow{2}{*}{111}  & Energy & 0.28 & 0.22 &  \multirow{2}{*}{4/30}\\
                         &                           &                       &                       & Forces & 0.27 & 0.19 & \\
      \hline
      \multirow{2}{*}{5} & \multirow{2}{*}{3}    & \multirow{2}{*}{12}   & \multirow{2}{*}{100}  & Energy & 0.41 & 0.33 &  \multirow{2}{*}{11/30} \\
                         &                           &                       &                       & Forces & 0.33 & 0.24 & \\
  \end{tabular}
  \caption{Test accuracies and MD instability for versions of PaiNN with various amounts of interaction layers.The cutoffs are measured in \si{\AA}, energy in \si{\kilo\cal\per\mol}, and forces in \si{\kilo\cal\per\mol\per\AA}. RMSE: Root Mean Squared Error, MAE: Mean Absolute Error. Data splits ($n_\mathrm{total}=85$k): $n_\mathrm{train}=0.85\times n_\mathrm{total}$, $n_\mathrm{val}=0.1\times n_\mathrm{total}$, $n_\mathrm{test}=0.05\times n_\mathrm{total}$.}
  \label{table:md_stability}
\end{table}

To show how the explainability framework in this study can be used to identify models which use chemically implausible prediction strategies, we trained 5 versions of PaiNN with different hyperparameters. The goal was to create a spectrum of models which range between common hyperparameters~\cite{schutt2018schnet,klicpera2020directional} to rather extreme hyperparameters which we anticipated to lead to models with chemically implausible representations. We varied the amount of interaction layers $L$ of PaiNN between $1$ and $5$ and adjusted the cutoff $c$ such that $L \cdot c = 15$~\AA, which is more than enough to cover the entire length of Ac-Ala3-NHMe even in its unfolded state. The range of networks, with ``unreasonable'' parameterizations at both ends, was chosen purely for didactic purposes: All networks were able to achieve a very low test error (see paragraph below) and interesting behavior can be observed at both ends of the spectrum.

For instance, it was known from previous studies~\cite{schutt2018schnet,klicpera2020directional} that cutoff lengths around 5 \AA~are well suited for MLFFs, and going significantly below 5~\AA, as we did here, impedes performance, but it is not entirely clear \textit{why} such short cutoff lengths do not work well (see section~\ref{sec:results:many-bodyness}).

Additionally to the number of interaction layers and the cutoff length, the number of radial basis functions and the embedding sizes were adjusted to keep the five networks comparable (see Table~\ref{table:md_stability}): The number of radial basis functions were varied such that their spacing along the distance between atoms is the same for all models. The embedding size was varied such that the total number of parameters is roughly equal. The 1-layer network is an exception, we set its embedding size to a larger value to keep its generalization error similar to the other networks. While it was not possible to keep the final generalization errors exactly equal (see Table~\ref{table:md_stability}), we note that all errors were far below what is generally considered ``chemically accurate'' (1 kcal/mol).

\bgroup
\renewcommand{\arraystretch}{1.2} % increase row spacing
\setlength{\tabcolsep}{4pt} % increase column spacing
\begin{table*}
  \rowcolors{4}{gray!25}{white}
  \begin{tabular}{l l l c  c  c  c }
    & & & \multicolumn{3}{c}{Interaction range measures} &  \\
    \cmidrule(lr){4-6}
    Model & Property & Data & $\lambda^\text{prob}_{0.001}$ (eq. \ref{eq:interaction_range_human_interpretable})  & $ \lambda^\text{pow}_1$ (eq. \ref{eq:generalized_expectation}) & $\lambda^\text{pow}_4$ (eq. \ref{eq:generalized_expectation}) &  \text{Many-bodyness} (eq. \ref{eq:manybodyness}) \\
    \hline \\
    3L SchNet           &  Energy  &  QM9                & 4.14 & 1.62 & 2.75 & 0.85  \\
3L SchNet           &  Dipole  &  QM9                & 6.34 & 2.64 & 3.37 & 1.40  \\
3L SchNet           &  HOMO    &  QM9                & 7.04 & 3.10 & 3.93 & 0.87  \\
3L SchNet           &  LUMO    &  QM9                & 7.01 & 3.03 & 3.76 & 0.95  \\
3L PaiNN            &  Energy  &  QM9                & 3.88 & 1.64 & 2.44 & 0.92  \\
3L PaiNN            &  Dipole  &  QM9                & 4.18 & 1.68 & 2.70 & 1.10  \\
3L PaiNN            &  HOMO    &  QM9                & 6.97 & 2.56 & 3.43 & 0.93  \\
3L PaiNN            &  LUMO    &  QM9                & 6.96 & 2.63 & 3.56 & 1.10  \\
1L PaiNN            &  Energy  &  Ac-Ala3-NHMe       & 8.55 & 2.63 & 4.41 & 0.70  \\
2L PaiNN            &  Energy  &  Ac-Ala3-NHMe       & 5.57 & 2.18 & 3.17 & 0.54  \\
3L PaiNN            &  Energy  &  Ac-Ala3-NHMe       & 4.03 & 2.26 & 2.98 & 0.80  \\
4L PaiNN            &  Energy  &  Ac-Ala3-NHMe       & 2.92 & 1.79 & 2.50 & 1.00  \\
5L PaiNN            &  Energy  &  Ac-Ala3-NHMe       & 2.57 & 1.71 & 2.29 & 1.70  \\

  \end{tabular}
  \caption{Interaction range and many-bodyness statistics. For the experiments with QM9 data, networks were trained and evaluated on the indicated property. For the experiments with Ac-Ala3-NHMe, networks were trained on energies and forces, and evaluated on the energies. For the interaction range, the thresholded range (eq. \ref{eq:interaction_range_human_interpretable}) with $p_\text{min}=0.001$ is displayed, and additionally the first and fourth generalized expectation of the  walk-length distribution (eq. \ref{eq:generalized_expectation} with $a=1$ and $a=4$). The many-bodyness has been evaluated with eq. \ref{eq:manybodyness}. For an extended table, see the supplement Table~\ref{table:all_statistics_full}.}
\label{table:all_statistics_short}
\end{table*}
\egroup

We conducted 30 MD simulations of Ac-Ala3-NHMe, with three simulations for each of 10 different starting configurations. The time-step of the integrator was $0.5$ \si{\femto\second} and the MD trajectories were run for 1 \si{\nano\second}, totaling 2 million time-steps. The simulations were performed in the canonical (NVT) ensemble at \SI{500}{\kelvin} with a time constant of \SI{5}{\femto\second}. An MD trajectory was considered unstable if the potential energy of the molecule went outside the range $-200$ to $200$ \si{\kilo\cal\per\mol}. In practice, this typically means that an atom of the molecule dissociated, which leads to an abrupt change in the energy. Table \ref{table:md_stability} shows how many MD trajectories per network were unstable.

It can be seen that the 1- and 5-layer networks were unstable, whereas the 2-, 3- and 4-layer networks were mostly stable. Note that the differences in force RMSE (Root Mean Squared Error) between the stable and unstable networks were negligible, with the most unstable network (1 layer) having one of the lower force errors. This indicates that the test error is not a good measure of how well a network will actually generalize, a finding that we replicate from other studies~\cite{stocker2022robust, fu2022forces,frank2024euclidean}. In the following sections, we will relate the interaction range and many-bodyness obtained from the adapted explainability framework introduced in this paper to the MD stability of these networks.

\subsection{Chemical principle 3: Interaction strength decreases polynomially with distance}
\label{sec:results:interaction_strength}

It is generally expected that the interaction strength between atoms beyond covalent bonds decreases with distance. However, there is no universal functional form to express this decrease. For example, the Coulomb force decreases with the inverse of the squared distance $r^{-2}$. Due to electric field screening effects, the effective decrease is typically much more rapid. In the Lennard-Jones potential, London dispersion forces decrease with $r^{-7}$ (due to the $r^{-6}$ term in the potential). What most decay laws have in common is that the decrease is proportional to a polynomial of the distance. We therefore expected to find that the interaction strength as seen by MLFFs would also decrease polynomially.

We tested this hypothesis using the 1-5-layer networks introduced above. Figure~\ref{fig:decay_per_layer}B presents exemplary interaction matrices for a single conformation of the tetrapeptide, alongside a matrix of pairwise atomic distances. These interaction matrices display the interaction strengths between atom pairs for the five different models, enabling qualitative comparison. In the 1-layer model, the interaction strengths are nearly uniformly distributed across all distances. As the number of layers increases, the interaction matrices become more diverse, indicating stronger interactions between atoms in close proximity and diminishing interaction strengths as the distance grows. This qualitative observation is quantitatively supported by statistical analyses across multiple samples, as shown in Figure~\ref{fig:decay_per_layer}D. It shows the relationship between the interatomic distance and the interaction strength. We first note that all networks show some decay with distance, but the degree with which the strength decays differs considerably. The 1-layer network plateaus after around \SI{5}{\AA}, which is chemically implausible and an indication for the poor MD stability of this network.

While the interaction strength decays in all of the four other networks, none of them exhibit a power law decay. In all cases, a decay modeled with an exponential is a better fit (see the helper lines in the plots indicating the best fit of an exponential curve). With each added interaction layer, the decay is faster. However, it is not immediately clear from this why the 5-layer network is unstable in MD trajectories, whereas the 4L network is mostly stable. We return to this question in section~\ref{sec:results:many-bodyness}.

In GNNs, the number of walks between two atoms decreases exponentially with distance. For an approximate formula for this decrease, see Appendix \ref{sec:number_of_walks}. What this means is that an exponential decrease of the interaction strength is ``baked in'' to GNNs, as long as they have a cutoff which is shorter than the length of the atomistic system that they operate on. Such an architectural constraint is also called an inductive bias in ML literature.

We note that one possible application of the framework introduced in this study is to guide the development of new MLFFs. For instance, now that we have a method to measure the interaction strength decay with distance, we can design alternative architectures and see what kind of decay they lead to. As an exemplary first step, in the next section we evaluate what happens in the absence of a cutoff.

\textbf{Interaction strength decreases without a cutoff.}
Our finding that the typical GNN architecture enforces an exponential decay of the interaction strength (see Section~\ref{sec:number_of_walks}) raises the question what would happen if the cutoff effects were removed. In the 1-layer network discussed in Section \ref{sec:results:interaction_strength}, we observed that it does not learn a consistent decay of the interaction strength and plateaus after about $5$\ \AA. However, since the model has only one interaction layer, the exact effect remains unclear, as GNNs are usually trained with at three layers or more.

Therefore, in order to test whether the failure to learn a decay after 5\ \AA\ is an isolated issue of having only one interaction layer, we trained PaiNNs with 3 interaction layers, with a cutoff length of $15$\ \AA, which is longer than the maximum length of the molecule. To further ``free'' the network from range constraints, we also removed the cosine cutoff, which is applied in many GNN-MLFF architectures and forces a cosine-shaped decay of the message features towards the end of the cutoff distance.

\begin{figure}
    \centering
    \includegraphics[width=0.5\textwidth]{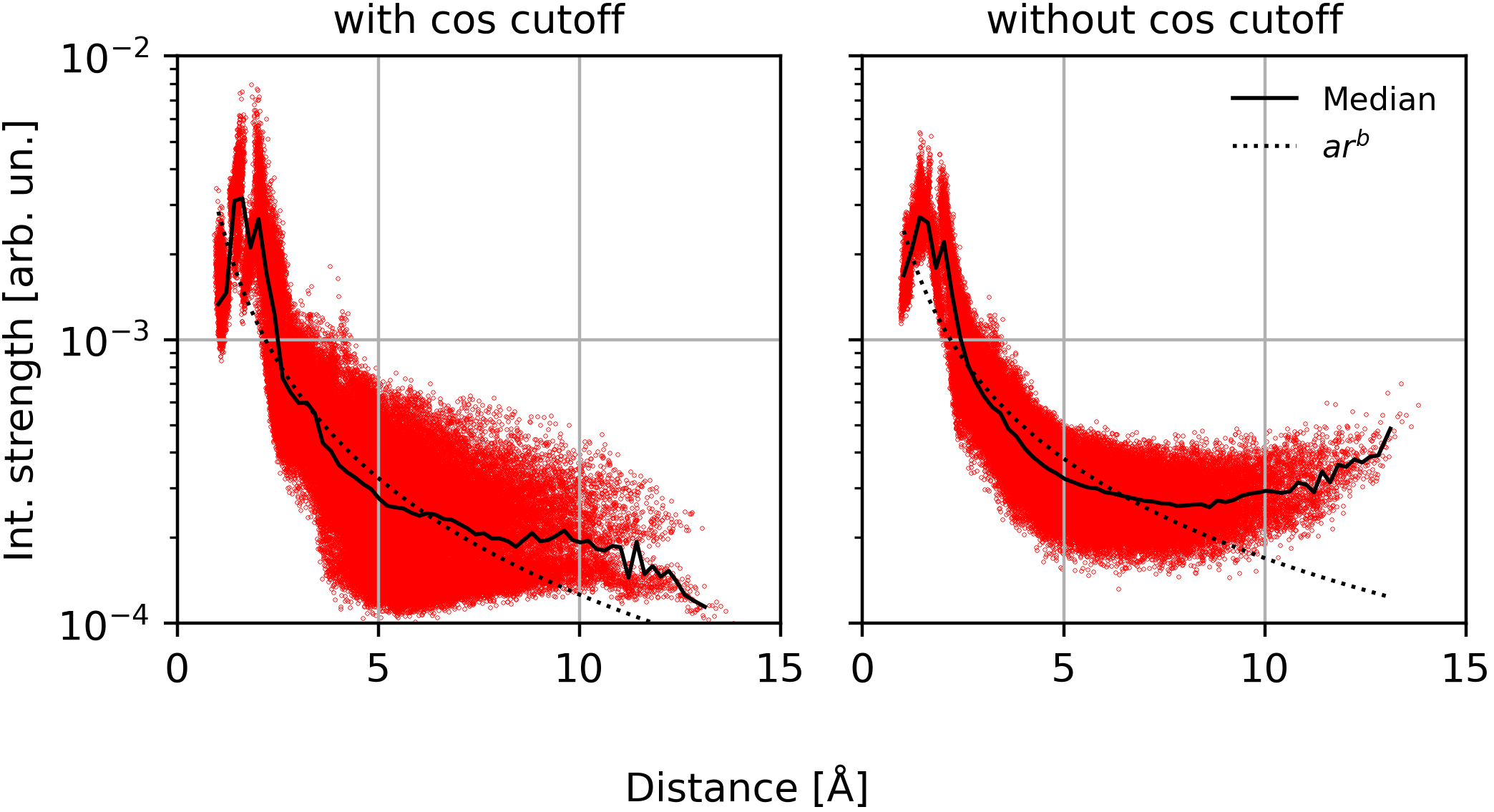}
    \caption{3 layer PaiNNs on Ac-Ala3-NHMe, with a cutoff length of 15\ \AA. This length is more than the maximum length of the molecule, so each atom in each layer of the GNN ``sees'' all other atoms directly. Left: with a cosine cutoff function; Right: without any cutoff function. The forces RMSE was $0.163$ and $0.159$ \si{\kilo\cal\per\mol\per\AA} respectively, so both networks had an almost equal validation error. The MD instability was 1/30 with cosine cutoff function, and 24/30 without. The dotted line is the best fit of a monomial function to the data.}
    \label{fig:3L15A_decay}
\end{figure}

The results show that in the absence of a cutoff and even without a cosine cutoff function applied, the model does not learn a chemically appropriate decay of the interaction strength (Figure~\ref{fig:3L15A_decay}, right). In fact, not only does the interaction strength not decay, but it increases again at higher distances. The tendency of the model training to increase the interaction strengths at higher distances is an effect which we observed throughout this study, and is explored in more depth in Section~\ref{sec:effect_of_training}. For the PaiNN model with a cosine-cutoff function applied, the decay of the interaction strength was more in line with the expected power law behavior. However, it does seem that the interaction strength almost plateaus before the influence of the applied cosine function reduces it further towards zero.

To investigate whether this chemical implausibility of the interaction strength reveals weaknesses in the learned representations, we performed the same MD-stability tests as described in Section~\ref{sec:md_stability}. Note that the forces RMSE of both networks were similar, at $0.163$ and $0.159$ \si{\kilo\cal\per\mol\per\AA}, respectively.
Despite this excellent validation error, the MD stability differed drastically. 29 out of 30 MD trajectories of the model with the cutoff were stable, compared to only 6 out of 30 without it. This indicates that networks that appear to be chemically implausible based on our analysis, do in fact extrapolate badly to new data, even if they seem indistinguishable from well functioning networks based solely on validation error.

\subsection{Chemical principle 4: Many-bodyness}
\label{sec:results:many-bodyness}

\begin{figure*}
    \centering
    \includegraphics[width=0.999\textwidth]{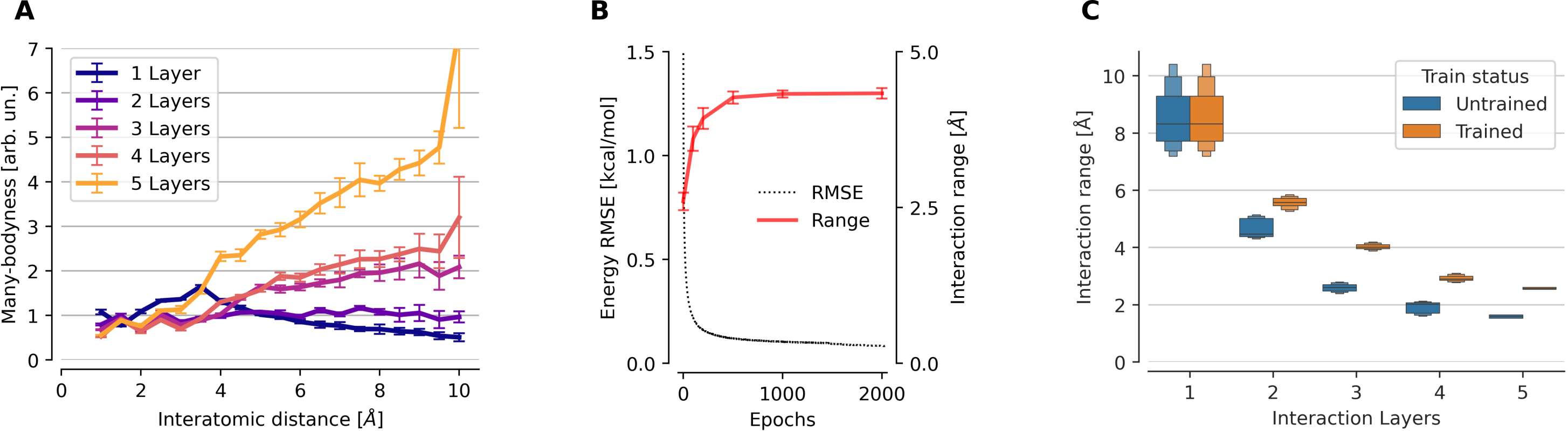}
    \caption{\textbf{(A)}: The many-bodyness $\gamma$ measured on several bins along the atom-pair distance. Note that the measure of the many-bodyness (eq.~\ref{eq:manybodyness}) is logarithmic (base 10), i.e.,\ a value of 1 unit of many-bodyness indicates that the lowest to the highest interaction strength in a bin differs by a factor of 10.  Networks: 1-5-layer PaiNNs trained and evaluated on Ac-Ala3-NHMe. \textbf{(B)}: Evolution of the energy RMSE and the mean interaction range (eq.~\ref{eq:interaction_range_human_interpretable}) during training. Error bars represent standard deviation. The 3-layer PaiNN architecture was trained on Ac-Ala3-NHMe. \textbf{(C)}: Interaction range (eq.~\ref{eq:interaction_range_human_interpretable}) for untrained and trained (on Ac-Ala3-NHMe) variants of the PaiNNs with 1-5 interaction layers.}
    \label{fig:manybodyness_and_range_by_training}
\end{figure*}

We defined the many-bodyness of atomic interactions as the base-10 log ratio of the strongest to the weakest interaction strength for atom pairs at the same distance (section~\ref{methods:many-bodyness}). The expectation is that atomic interactions are influenced by other atoms in the neighborhood, modulating the interaction.

We contrast the expectation of many-bodyness of the interaction as seen by MLFFs with classical force fields. The 2-body terms in classical force fields are fully isotropic, as the effect of other atoms can not be taken into consideration by definition. 3- and 4-body terms do take other atoms into consideration and would lead to the possibility of at least some many-bodyness even in classical force fields. However, 3-, 4- and higher order terms are typically only applied to (chains of) covalently bonded atoms. This means that atom pairs at higher distances will experience strictly isotropic interactions in classical force fields.

As seen already in section~\ref{sec:results:interaction_strength} (Figure~\ref{fig:decay_per_layer}DE), the interaction strengths as seen by MLFFs differ significantly at the same distance. At first thought, one may assume that the 1-layer network has no many-bodyness, because it considers only 2-body terms. However, this is not true: In the 1-layer network, each atom receives input from all other atoms in the molecule, and then integrates all of these ``messages'' into its final prediction. A 1-layer GNN is therefore not equivalent to 2-body terms.

We also compared the many-bodyness across interaction distances (Figure~\ref{fig:manybodyness_and_range_by_training}A). It is known that local interactions should be influenced by the other atoms in the neighborhood, which is therefore what we expected to find in GNNs.
We found that the many-bodyness grows with increasing distance in the 3-, 4- and 5-layer networks. For the 2-layer network, the many-bodyness stays roughly constant and increases only slightly for larger interatomic distances, and for the 1-layer network it decreases with distance.

The many-bodyness of the 5-layer network seems excessively high: It reaches a value above 4 even at relatively short distances (below $8$\ \AA), which means that the interaction strengths differ by a factor of more than $10,000$. We hypothesize that this high many-bodyness is not physical and is an indicator as to why the MD-trajectories that were run with this network are often unstable.

\subsection{How does training a GNN change the interaction range?}
\label{sec:effect_of_training}
Using our measure of the interaction range can not only be used on the fully trained network. Instead, the evolution of these measures can be tracked throughout the training of the GNN. Doing this analysis uncovers that the interaction range is increased significantly during training, but only after the error on the test set already improved significantly (Figure~\ref{fig:manybodyness_and_range_by_training}B). As the validation error approaches a plateau, the interaction range keeps increasing. We hypothesize that this is because the error can initially be reduced by taking into consideration only the immediate surrounding of each atom, whereas to remove the last remaining bits of error, a wider context needs to be considered.

A noteworthy finding is that training of the model increases the interaction range in all cases, even when the range in the untrained model (i.e., a model with randomly initialized weights) starts out higher than what is likely physically appropriate, as is the case in the 1-layer network. Figure~\ref{fig:manybodyness_and_range_by_training}C shows this effect: For each of the 1- up to 5-layer PaiNNs, the trained variant has a longer interaction range than the untrained one. For the 1-layer variant, the intuitive interaction range measure which is shown in this figure does not distinguish between trained and untrained, because in both cases, the threshold for the range cutoff is higher than the length of the molecule. The interaction range measure based on the fourth moment of the walk distribution (Table \ref{table:all_statistics_short}) however shows that the range of the trained 1-layer network is indeed significantly higher.

\section{Discussion}
MLFFs methods for computational chemistry have recently become highly popular since they are considered a useful compromise between classical force fields (quick, less accurate) and first-principles electronic structure calculations (slow, more accurate). Recently a zoo of different kernel and deep learning models (e.g., \cite{noe2020machine,NEURIPS2020_15231a7c,unke2021machine,keith2021combining,MD22,batzner2021se,unke2024biomolecular,frank2024euclidean}) have emerged, offering many possible MLFF modelling approaches. Throughout the use of such MLFFs, the community has been striving to gain a deeper understanding of their potential limitations. While for general applications in the sciences (e.g., \cite{klauschen-pathology24, binder2021morphological, eberle2020building, el2023explainability}), XAI methods are helpful in the quest of gaining insight into ML models, their use in theoretical chemistry has so far been rather limited (see e.g.\ \cite{schutt2017quantum,jose20drugdisc,mcclo19usingatt,ying2019gnnexplainer,luo2020parameterized,bonneau2024peering,gallegos_explainable_2024}).
Moreover, XAI methods have been used to debug models, to gain novel insights and to capture whether or not suspected/expected structures or knowledge are embodied in the respective ML architectures (see e.g.\ \cite{samek2021explaining}).

In this work we have analyzed how much two popular MLFFs (SchNet and PaiNN) reflect chemical principles after training. Such principles were typically embodied in the  GNNs: They were able to extract physical relationships from data just by learning to predict a set of energies and forces. At the same time, one further important property, namely that the interaction strength between atom pairs should decrease with a power law, was violated. Indeed, we showed theoretically and experimentally that a fundamental limitation of current GNN architectures is that the interaction strength decreases exponentially. Moreover, in all practically relevant cases, a cutoff parameter leads to exponential decay of the interaction strength. An exponential decay  is not in agreement with physicochemical principles. Especially when imposing a cutoff distance of $4-5 \ \text{\AA}$, as is common in state-of-the-art MLFFs, distances above $10 \text{\AA}$ can  barely be modelled. This finding can be taken as guidance to design improved GNNs that fulfill power-law properties (or interaction distributions as proposed in Ref.~\citenum{gori2023second})  and can in this manner closer reflect the underlying chemistry and physics.

A somewhat troubling finding was that several different GNNs (e.g., the model variants using too few or too many layers or unsuitable cutoffs), while having a very low test-set error after training), differed significantly in their learned prediction strategy. This means that a model does not necessarily have to reflect the known underlying principles well in order to yield a good prediction quality. It had been shown previously~\cite{stocker2022robust, fu2022forces,frank2024euclidean} that the test-set error is not necessarily indicative of MD stability---a finding clearly replicated in this study. However, with the XAI-based analysis proposed, we can obtain deeper insights as we can show that models which deviate too far from underlying physical principles will produce unstable MD trajectories, despite their low test set error.

Our findings suggest that ML models applied to chemical systems can still benefit from several improvements that encode the physicochemical principles discussed in this work. This could lead to enhanced transferability in compositional and structural chemical spaces as well as scalability in terms of system size.

These results show a tangible benefit of analyzing MLFFs with explainability methods. Specifically, they confirm that MLFFs can indeed learn the fundamental physical and chemical principles as expected, which allows a more confident transition of MLFFs from exploratory research to real-world applications.

\section*{Acknowledgement}
This work was in part supported by the German Ministry for Education and Research (BMBF) under Grants 01IS14013A-E, 01GQ1115, 01GQ0850, 01IS18025A, 031L0207D, and 01IS18037A, and by BASLEARN---TU Berlin/BASF Joint Laboratory, co-financed by TU Berlin and BASF SE. K.R.M.\ was partly supported by the Institute of Information \& Communications Technology Planning \& Evaluation (IITP) grants funded by the Korea government (MSIT) (No. 2019-0-00079, Artificial Intelligence Graduate School Program, Korea University and No. 2022-0-00984, Development of Artificial Intelligence Technology for Personalized Plug-and-Play Explanation and Verification of Explanation). A.K. acknowledges financial support from the Luxembourg National Research Fund (FNR AFR Ph.D. Grant 15720828).
AT was funded by the European Research Council (ERC Advanced Grant FITMOL).
Correspondence to AT and KRM.

\bibliography{references.bib}

\clearpage
\renewcommand{\thesection}{S\arabic{section}}
\renewcommand{\thetable}{S\arabic{table}}
\renewcommand{\thefigure}{S\arabic{figure}}
\setcounter{figure}{0}
\onecolumngrid
\appendix
\section{Appendix}
\subsection{Interaction strength, Interaction range and Many-bodyness computation}
\label{sec:appendix:methods}

\begin{algorithm}
\SetAlgoLined
\DontPrintSemicolon
\KwData{Graph $\mathcal{G}$; model; walk distance function $g$}
\KwResult{Interaction range, interaction strengths, and many-bodyness}
\tcp{Model forward pass (prediction)}
$\hat{y} \leftarrow \text{model}(\mathcal{G})$\;
\BlankLine
\tcp{Relevance backward pass with gnnlrp library based on Schnake et al. 2022}
$(\{\mathcal{W}\}, \{\mathcal{R}_{\mathcal{W}}\}) \leftarrow \text{gnnlrp}(\mathcal{G}, \text{model}, \hat{y})$\;
\For{$\mathcal{W} \in \{\mathcal{W}\}$}{
  $\mathbb{P}_{\mathcal{W}} \leftarrow \frac{|\mathcal{R}_{\mathcal{W}}|}{\sum_{\mathcal{W}'}|\mathcal{R}_{\mathcal{W}'}|}$\;
}
\tcp{Fourth generalized expectation of the walk distribution}
  $\lambda \leftarrow \big[ \sum_{\mathcal{W}}\mathbb{P}_{\mathcal{W}} d(\mathcal{W}^4) \big]^{\tfrac{1}{4}}$\;
\BlankLine
\tcp{Interaction strength computation}
\ForEach{atom pair $(i, j)$ in $\text{atom\_pairs}(\mathcal{G})$}{
  \tcp{Get the set of all walks in which atoms $i$ and $j$ are included}
  $\mathcal{S}_{ij} = \{\mathcal{W} | i \in \mathcal{W} \wedge j \in \mathcal{W} \}$\;
  $s_{ij} \leftarrow \sum_{\mathcal{W} \in \mathcal{S}_{ij}} \mathbb{P}_{\mathcal{W}}$\;
  % Atom pair interaction strengths
}
\tcp{Many-bodyness computation}
\tcp{Binning of interaction strengths based on the interatomic distance}
$\{s\}_R \leftarrow \{s_{ij} | \, || \bm{r}_{i} - \bm{r}_{j} || \approx R \}$\;
$\gamma \leftarrow \frac{1}{R_{max}} \int_0^{R_{\text{max}}} \log_{10} \Bigg(\frac{P_{100}(\{s\}_R)}{P_{10}(\{s\}_R)}\Bigg) \, dR$\;
\KwRet{\text{Interaction range} $\lambda$, \text{Many-bodyness} $\gamma$}
\caption{Algorithm for computing interaction range and many-bodyness}
\label{algorithm_xai}
\end{algorithm}

\subsection{Interaction strengths per element and distance}
\label{sec:appendix:star_plot}
\begin{figure}[h]
    \centering
    \includegraphics[width=0.5\textwidth]{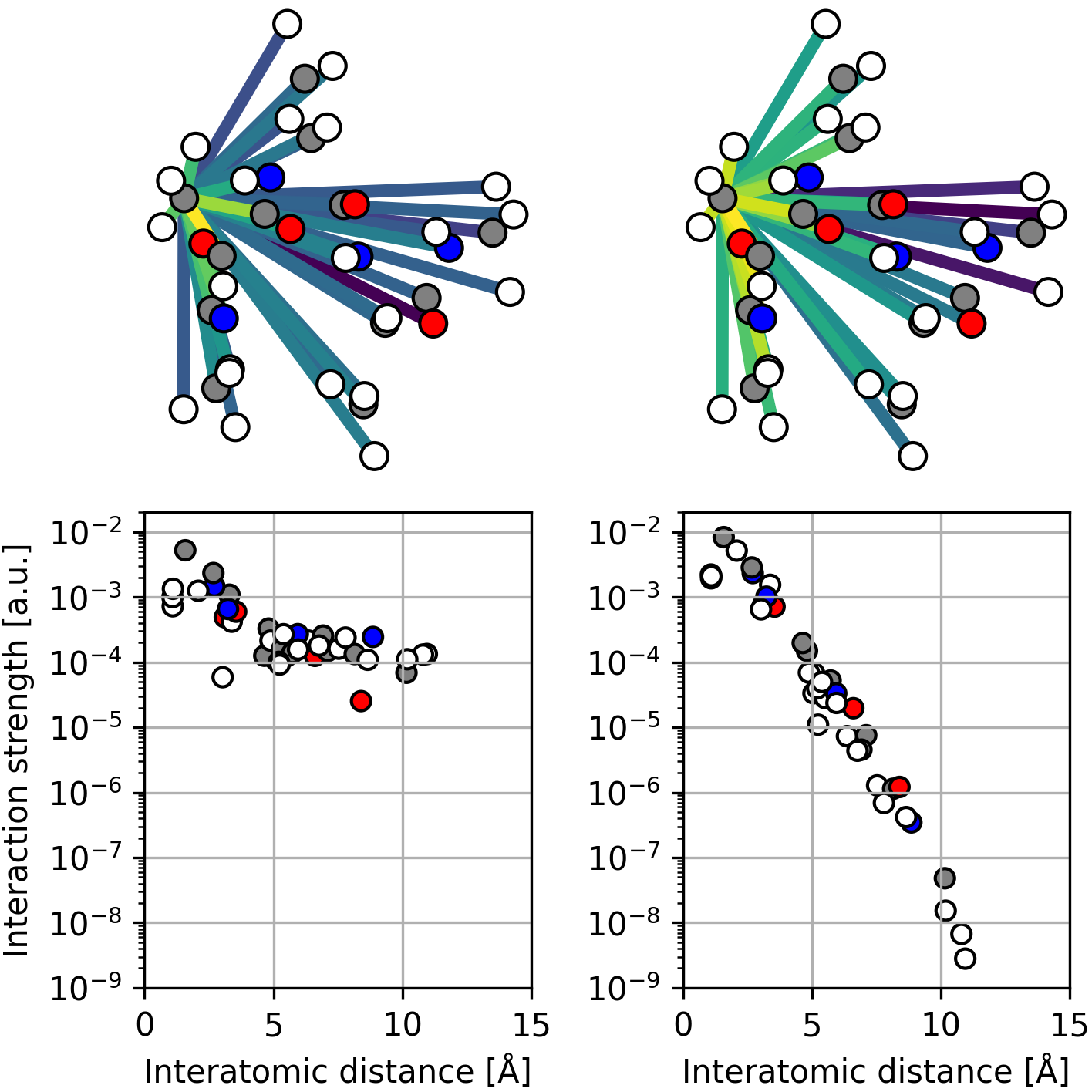}
    \caption{Upper row: interaction strengths between a given atom and all other atoms in a conformation of AcAla3-NHMe are indicated by the color of the connecting lines. Each plot in the upper row corresponds to a plot in the the lower row, where the same atom pair interaction strengths are visualized. The colors indicate the element of the partner atom: White: Hydrogen, Grey: Carbon, Blue: Nitrogen, Red: Oxygen.}
    \label{fig:star_plot}
\end{figure}

Due to space constraints, the figures in the main text only show how the interaction strength changes with interatomic distance for all atom pairs. Another interesting aspect is what role the elements play. Figure~\ref{fig:star_plot} indicates shows this relationship. At a first glance it seems that the type of element is not relevant, however keep in mind that the y-axis has a log-scale, and even smaller distances along the y-axis can be significant.\\
What complicates this analysis is that hydrogen is mostly found at the outer part of the molecule, where interaction strengths are generally weaker than in the center. This can partially explain the relatively weaker interactions with hydrogen atoms.

\subsection{Number of walks decreases exponentially}
\label{sec:number_of_walks}
Since our definition of the atom interaction strength involves a sum over walk relevances, the number of walks plays a central role for the interaction strength. The number of walks between two atoms can be approximated in terms of their distance and the number of interaction layers of the GNN.\\
We denote with $|{\mathcal{S}}_{ij}|$ the number of walks between atoms $i$ and $j$. It can be approximated as:
\begin{equation}
    |{\mathcal{S}}_{ij}| =
\begin{cases}
\left(\frac{4\pi}{3} c^3 \rho\right)^{L} & \text{if } d \leq c \\
\left(\frac{4\pi}{3} c^3 \rho\right)^{(L - \frac{d}{c})} & \text{if } c < d \leq Lc \\
0 & \text{if } d > Lc
\end{cases}
\end{equation}
where $d$ is the distance between two atoms, $c$ is the cutoff length, $L$ is the number of message passing layers and $\rho$ is the density of atoms. The term $\frac{4\pi}{3} c^3 \rho$ can be thought of as the average number of neighbors.\\
One can see that there's a discontinuity at the cutoff distance. The discontinuity can be explained by the fact that above the cutoff, at least two "hops" are needed to connect two atoms, which abruptly constrains the total number of possible walks between two atoms. Interestingly, when considering the atom interaction strengths, this discontinuity is barely visible anymore (see Figure~\ref{fig:decay_per_layer}). This smoothing out of the discontinuity is due to the cosine cutoff of the GNN filters.\\
\begin{figure}[h]
    \centering
    \includegraphics[width=0.4\textwidth]{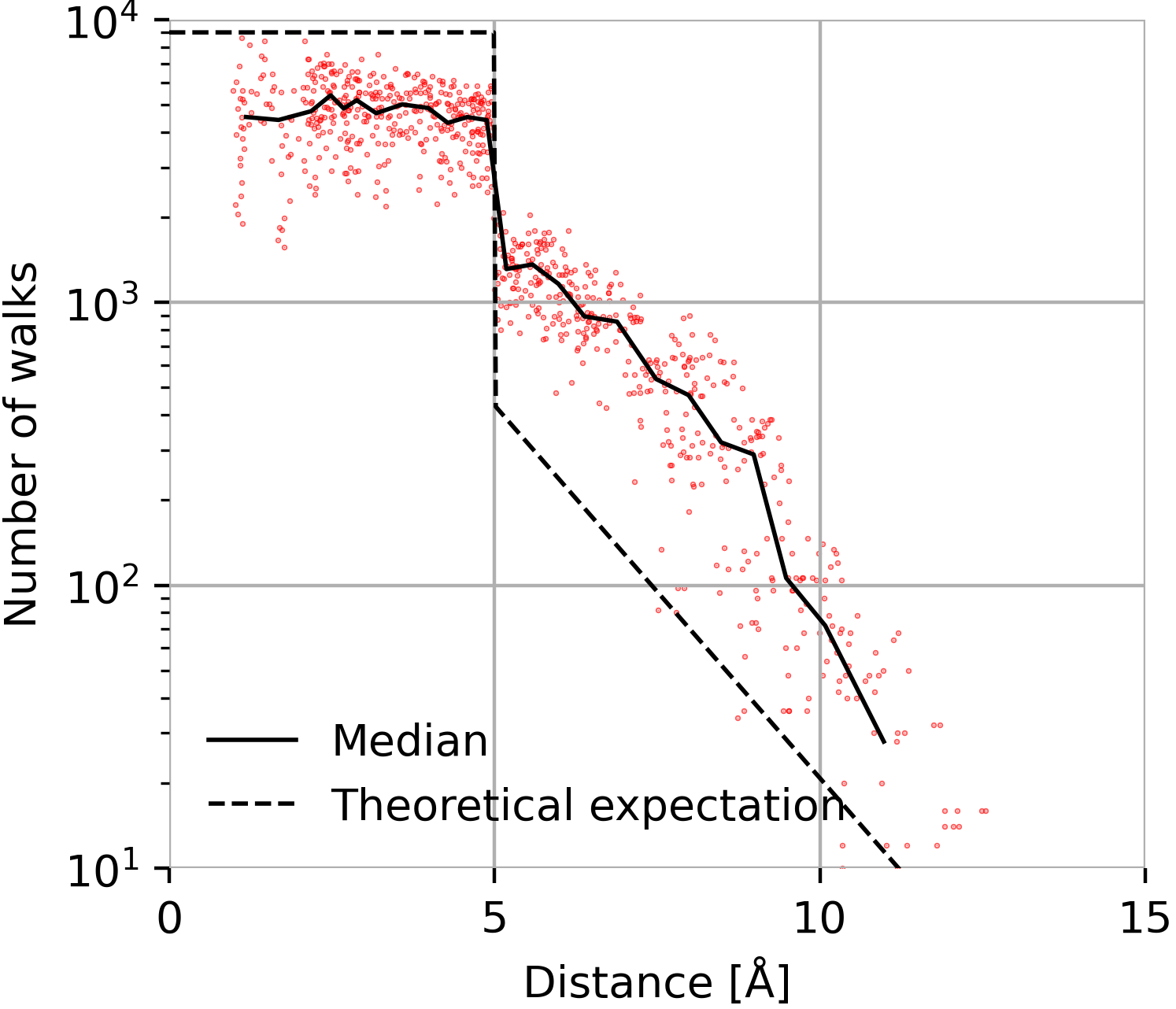}
    \caption{Theoretically expected and empirically found number of walks for each pair of atoms. The GNN is a PaiNN with three interaction layers and a cutoff of 5 \AA. The molecule is AcAla3-NHMe. Each point represents an atom pair.}
    \label{app:fig:theoretical_number_of_walks}
\end{figure}
The predicted number based on this formula aligns qualitatively with the empirically found number of walks (Figure~\ref{app:fig:theoretical_number_of_walks}). Importantly, the slope of the prediction and the empirical median line are similar. This shows that both theoretically and empirically, the number of walks decreases expentially with the interatomic distance. This relationship is fundamentally the same for all GNNs with a cutoff and indicates an inductive bias of this architecture.

\subsection{Further results statistics}
Here, we show the full results table, an extension of Table~\ref{table:all_statistics_short} from the main text. In the full table, we sometimes evaluate the statistics on a different molecule than what we trained on. This is to make the results more comparable: except for the QM9 results, we always evaluate on the AcAla3-NHMe molecule. In some cases, we want to find out the effect of training on different molecules, and train on the molecule indicated in the table row. For a discussion of training on various other molecules, see \ref{sec:other_molecules_training}.
\bgroup
\renewcommand{\arraystretch}{1.2} % increase row spacing
\setlength{\tabcolsep}{3pt} % increase column spacing
\begin{table*}
  \rowcolors{4}{gray!25}{white}
  \begin{tabular}{l l l c  c  c  c }
    & & & \multicolumn{3}{c}{Interaction range measures} &  \\
    \cmidrule(lr){4-6}
    Model & Property & Data & $\lambda^\text{prob}_{0.001}$ (eq. \ref{eq:interaction_range_human_interpretable})  & $ \lambda^\text{pow}_1$ (eq. \ref{eq:generalized_expectation}) & $\lambda^\text{pow}_4$ (eq. \ref{eq:generalized_expectation}) &  \text{Many-bodyness} (eq. \ref{eq:manybodyness}) \\
    \hline \\
    3L SchNet           &  Energy  &  QM9                & 4.14 & 1.62 & 2.75 & 0.85  \\
3L SchNet           &  Dipole  &  QM9                & 6.34 & 2.64 & 3.37 & 1.40  \\
3L SchNet           &  HOMO    &  QM9                & 7.04 & 3.10 & 3.93 & 0.87  \\
3L SchNet           &  LUMO    &  QM9                & 7.01 & 3.03 & 3.76 & 0.95  \\
3L PaiNN            &  Energy  &  QM9                & 3.88 & 1.64 & 2.44 & 0.92  \\
3L PaiNN            &  Dipole  &  QM9                & 4.18 & 1.68 & 2.70 & 1.10  \\
3L PaiNN            &  HOMO    &  QM9                & 6.97 & 2.56 & 3.43 & 0.93  \\
3L PaiNN            &  LUMO    &  QM9                & 6.96 & 2.63 & 3.56 & 1.10  \\
1L PaiNN            &  Energy  &  Ac-Ala3-NHMe       & 8.55 & 2.63 & 4.41 & 0.70  \\
2L PaiNN            &  Energy  &  Ac-Ala3-NHMe       & 5.57 & 2.18 & 3.17 & 0.54  \\
3L PaiNN            &  Energy  &  Ac-Ala3-NHMe       & 4.03 & 2.26 & 2.98 & 0.80  \\
4L PaiNN            &  Energy  &  Ac-Ala3-NHMe       & 2.92 & 1.79 & 2.50 & 1.00  \\
5L PaiNN            &  Energy  &  Ac-Ala3-NHMe       & 2.57 & 1.71 & 2.29 & 1.70  \\
1L PaiNN untrained  &  Energy  &  Ac-Ala3-NHMe       & 8.55 & 1.15 & 3.69 & 0.46  \\
2L PaiNN untrained  &  Energy  &  Ac-Ala3-NHMe       & 4.68 & 1.05 & 2.99 & 0.40  \\
3L PaiNN untrained  &  Energy  &  Ac-Ala3-NHMe       & 2.60 & 0.52 & 1.89 & 0.78  \\
4L PaiNN untrained  &  Energy  &  Ac-Ala3-NHMe       & 1.94 & 0.20 & 1.27 & 1.30  \\
5L PaiNN untrained  &  Energy  &  Ac-Ala3-NHMe       & 1.63 & 0.087 & 0.88 & 2.30  \\
3L SchNet           &  Energy  &  Ac-Ala3-NHMe       & 4.54 & 2.62 & 3.34 & 0.72  \\
3L PaiNN            &  Energy  &  ATAT-CGCG          & 4.61 & 2.60 & 3.41 & 0.76  \\
3L PaiNN            &  Energy  &  MD22 (exc. tubes)  & 3.73 & 1.95 & 2.90 & 0.76  \\

  \end{tabular}
  \caption{Interaction range and many-bodyness statistics. For the experiments with QM9 data, networks were trained and evaluated on the indicated property. For the experiments with Ac-Ala3-NHMe, networks were trained on energies and forces, and evaluated on the energies. For the interaction range, the thresholded range (eq \ref{eq:interaction_range_human_interpretable}) with $p_\text{min}=0.001$ is displayed, and additionally the first and fourth generalized expectation of the  walk-length distribution (eq \ref{eq:generalized_expectation} with $a=1$ and $a=4$). The many-bodyness has been evaluated with eq \ref{eq:manybodyness}. The row titled "3L PaiNN trained on MD22" is for a PaiNN that has been trained on all molecules in the MD22 dataset (except the nanotubes).}
  \label{table:all_statistics_full}
\end{table*}

\subsection{Influence of training data on interaction range}
\label{sec:other_molecules_training}
\begin{figure*}[h]
    \centering
    \includegraphics[width=0.55\textwidth]{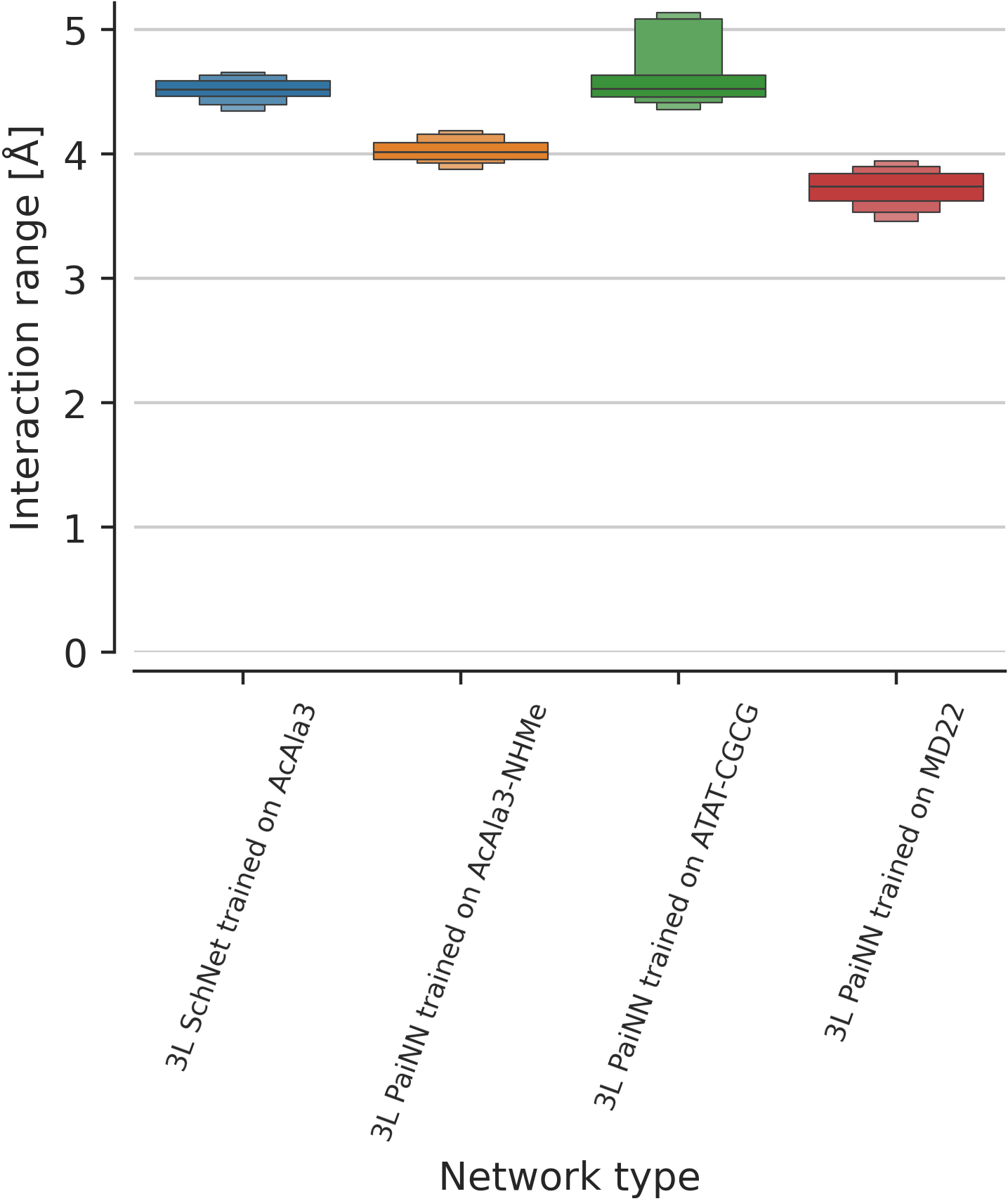}
    \caption{Interaction range for various networks. The interaction ranges have been evaluated on AcAla3-NHMe in each case, regardless of the molecule they were trained on.}
    \label{fig:boxplot_various_networks}
\end{figure*}
\begin{figure*}[h]
    \centering
    \includegraphics[width=0.55\textwidth]{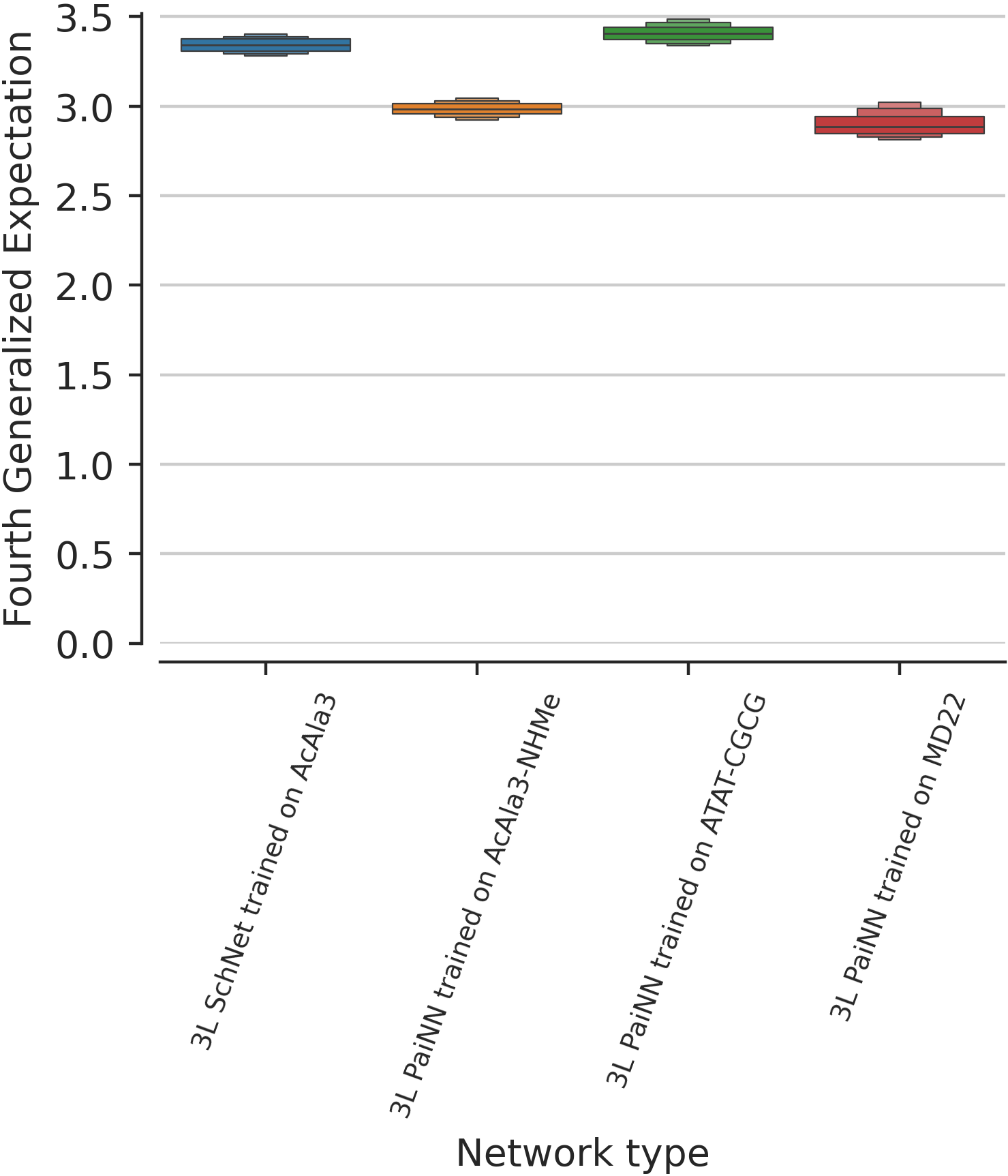}
    \caption{Fourth generalized expectation of the walk distribution (eq~\ref{eq:generalized_expectation}) for various networks. The interaction ranges have been evaluated on AcAla3-NHMe in each case, regardless of the molecule they were trained on.}
    \label{fig:boxplot_various_networks_fourth_moment}
\end{figure*}

We trained 3L PaiNNs on AcAla3-NHMe, ATAT-CGCG (also from MD22) and on all molecules in MD22 combined (except the nanotubes) to find out whether the type of molecule that was trained on has an effect on the interaction range. For comparison with another NN architecture, we also compare to a SchNet trained on AcAla3-NHMe. In the direct comparison between SchNet and PaiNN, there seems to be no difference (Figure~\ref{fig:boxplot_various_networks}). However, when using the more detailed but less interpretable range measure based on the fourth moment of the walk distribution, we see PaiNN has a somewhat lower interaction range than SchNet (Figure~\ref{fig:boxplot_various_networks_fourth_moment}), which is a trend that we already observed on the QM9 data (Figure~\ref{fig:boxplot-extensive-intensive}). Comparing the three PaiNNs trained on AcAla3-NHMe, ATAT-CGCG and all molecules in MD22 shows that the model trained on all molecules has a significantly lower interaction range, whereas the one trained on ATAT-CGCG has a higher range. We speculate that fitting the long-distance interactions on all molecules at the same time is too difficult because the long-distance interactions are heavily dependent on the type of molecule. The short-range interactions on the other hand are more similar between molecules and are also more important factors for predicting the energy/forces than the long range interactions. Therefore, the training focuses the model mostly on the short-range interactions, which leads to a lower interaction range. On the other hand, ATAT-CGCG is a larger system than AcAla3-NHMe, comprised of four molecules with long-range inter-molecular forces. Here, in line with this fact, the interaction range is the largest.\\

\subsection{Computational details: GNN Training and MD}
\label{sec:training_details}
All results throughout the paper were computed according to algorithm \ref{algorithm_xai}. The models were trained using SchNetPack \cite{schutt2019schnetpack,schutt2023schnetpack}, using the parameters in Table\ \ref{table:train_details} and Table\ \ref{table:layers_train_details}. 
\begin{table*}[]
\begin{tabular}{ll}
Parameter                & Value                \\
\hline
Batch size               & 100                  \\
Weight decay             & 0.01                 \\
Learning rate            & 0.0005               \\
N interaction layers     & 3                    \\
Atomic feature embedding & 128                  \\
Radial basis function    & Gaussian             \\
N radial basis functions & 20                   \\
Cutoff function          & Cosine               \\
Cutoff length [\AA]      & 5                    \\
\end{tabular}
\caption{Parameters used for training. For the 1-5 layer experiments, some parameters were adapted as in Table~\ref{table:layers_train_details}.}
\label{table:train_details}
\end{table*}

\begin{table*}[]
\begin{tabular}{llllll}
N interaction layers     & 1   & 2   & 3   & 4    & 5   \\
\hline
Atomic feature embedding & 512 & 157 & 128 & 111  & 100 \\
N radial basis functions & 60  & 30  & 20  & 15   & 12  \\
Cutoff length [\AA]         & 15  & 7.5 & 5   & 3.75 & 3  
\end{tabular}
\caption{When traning the 1-5 layer networks, it was tried to have them have roughly equal weights, while still maintaining comparable validation errors. The parameters not listed here are as in Table~\ref{table:train_details}.}
\label{table:layers_train_details}
\end{table*}

\end{document}